\documentclass{elsart}
\usepackage{graphics}

\begin{document}

\def\etal{{\it{}et~al.}}        
\def\eref#1{(\protect\ref{#1})}
\def\fref#1{\protect\ref{#1}}
\def\sref#1{\protect\ref{#1}}
\def\tref#1{\protect\ref{#1}}
\def\av#1{\langle#1\rangle}
\def\th#1{$#1^{\rm th}$}

%
\def\tcapt#1{\refstepcounter{table}\bigskip\hbox to \textwidth{%
       \hfil\vbox{\hsize=\captwidth\renewcommand{\baselinestretch}{1}\small
       {\sc Table \thetable}\quad#1}\hfil}\bigskip}

%
\newdimen\captwidth
\captwidth=13cm                                
\newdimen\sidecaptwidth
\sidecaptwidth=5cm                             
\newdimen\normalfigwidth
\normalfigwidth=\captwidth                     
\newdimen\smallfigwidth
\smallfigwidth=11cm                            
\newdimen\tinyfigwidth
\tinyfigwidth=4cm                              
\newdimen\realtinyfigwidth
\realtinyfigwidth=3cm                          
\newdimen\sidefigwidth
\sidefigwidth=7cm                              

\def\capt#1{\refstepcounter{figure}\bigskip\hbox to \textwidth{%
       \hfil\vbox{\hsize=\captwidth\renewcommand{\baselinestretch}{1}\small
       {\sc Figure \thefigure}\quad#1}\hfil}\bigskip}

\def\normalfigure#1{\hbox to\textwidth{%
       \hfil\resizebox{\normalfigwidth}{!}{\includegraphics{#1}}\hfil}}

\def\smallfigure#1{\hbox to\textwidth{%
       \hfil\resizebox{\smallfigwidth}{!}{\includegraphics{#1}}\hfil}}

\def\ninefigure#1#2#3#4#5#6#7#8#9{\hbox to \textwidth{%
       \hfil\hbox to\normalfigwidth{%
       \resizebox{\tinyfigwidth}{!}{\includegraphics{#1}}\hfil
       \resizebox{\tinyfigwidth}{!}{\includegraphics{#2}}\hfil
       \resizebox{\tinyfigwidth}{!}{\includegraphics{#3}}}\hfil}
       \bigskip
       \hbox to \textwidth{%
       \hfil\hbox to\normalfigwidth{%
       \resizebox{\tinyfigwidth}{!}{\includegraphics{#4}}\hfil
       \resizebox{\tinyfigwidth}{!}{\includegraphics{#5}}\hfil
       \resizebox{\tinyfigwidth}{!}{\includegraphics{#6}}}\hfil}
       \bigskip
       \hbox to \textwidth{%
       \hfil\hbox to\normalfigwidth{%
       \resizebox{\tinyfigwidth}{!}{\includegraphics{#7}}\hfil
       \resizebox{\tinyfigwidth}{!}{\includegraphics{#8}}\hfil
       \resizebox{\tinyfigwidth}{!}{\includegraphics{#9}}}\hfil}}

\def\fourfigure#1#2#3#4{\hbox to \textwidth{%
       \hfil\hbox to\normalfigwidth{%
       \resizebox{\realtinyfigwidth}{!}{\includegraphics{#1}}\hfil
       \resizebox{\realtinyfigwidth}{!}{\includegraphics{#2}}\hfil
       \resizebox{\realtinyfigwidth}{!}{\includegraphics{#3}}\hfil
       \resizebox{\realtinyfigwidth}{!}{\includegraphics{#4}}}\hfil}}

\def\sidefigure#1#2{\hbox to \textwidth{%
       \resizebox{\sidefigwidth}{!}{\includegraphics{#1}}\hfil
       \refstepcounter{figure}
       \vbox{\hsize=\sidecaptwidth\renewcommand{\baselinestretch}{1}
       \small\raggedright{\sc Figure \thefigure}\quad#2}}}

\date{19 March 1997}
\journal{Los Alamos preprint archives}

\begin{frontmatter}
\title{New Monte Carlo algorithms for classical spin systems}
\author{G. T. Barkema}
\address{HLRZ, Forschungszentrum J\"ulich, 52425 J\"ulich, Germany}
\author{M. E. J. Newman}
\address{Santa Fe Institute, 1399 Hyde Park Road, Santa Fe, NM 87501.
  U.S.A.}
\author{\ }
\address{{\rm To appear in {\sl Monte Carlo Methods in Chemical Physics},\\
  D. Ferguson, J. I. Siepmann, and D. G. Truhlar (eds.),\\
  Wiley, New York (1997).}}
\begin{abstract}
  We describe a number of recently developed cluster-flipping
  algorithms for the efficient simulation of classical spin models
  near their critical temperature.  These include the algorithms of
  Wolff, Swendsen and Wang, and Niedermeyer, as well as the limited
  cluster algorithm, the multigrid methods of Kandel and co-workers,
  and the invaded cluster algorithm.  We describe the application of
  these algorithms to Ising, Potts, and continuous spin models.
\end{abstract}
\end{frontmatter}

Monte Carlo simulations of classical spin systems such as the Ising model
can usually be performed simply and very efficiently using local update
algorithms such as the Metropolis or heat-bath algorithms.  However, in the
vicinity of a continuous phase transition, these algorithms display
dramatic critical slowing down, making them extremely poor tools for the
study of critical phenomena.  In the last ten years or so, a number of new
Monte Carlo algorithms making use of non-local update moves have been
developed to address this problem.  In this chapter we examine a number of
these algorithms in some detail, and discuss their strengths and
weaknesses.  The outline of the chapter is as follows.  In
Section~\sref{csd}, we briefly discuss the problem of critical slowing
down.  In Section~\sref{wolff} we describe the Wolff algorithm, which is
probably the most successful of the algorithms developed so far.  In
Section~\sref{further} we discuss a number of other algorithms, such as the
Swendsen-Wang algorithm, Niedermayer's algorithm, and the invaded cluster
algorithm.  Each of these algorithms is introduced first in the context of
the Ising model, but in Section~\sref{pottsetc} we discuss how they may be
generalized to systems with many valued spins such as Potts models or
continuous valued spins such as the classical XY and Heisenberg models.  In
Section~\sref{conclusions} we give our conclusions.

\section{Critical slowing down}
\label{csd}
The divergence of the correlation length $\xi$ of a classical spin system
as it approaches its critical temperature means that larger and larger
volumes of spins must be updated coherently in order to sample
statistically independent configurations.  In Monte Carlo simulations this
gives rise to a correlation time $\tau$ measured in Monte Carlo steps
per site which diverges with $\xi$ as
\begin{equation}
\tau \sim \xi^z,
\end{equation}
where $z$ is a dynamic exponent whose value depends on the update method.
For a system of finite dimension $L$ close to the critical temperature,
this means that the amount of CPU time required to generate an independent
lattice configuration increases as $L^{d+z}$, where $d$ is the
dimensionality of the lattice.  If we take the example of the Ising model
and make the assumption that the movement of domain walls under a dynamics
employing a local update move is diffusive, then the number of Monte Carlo
steps required to generate an independent spin configuration should scale
as $L^{d+2}$, and hence $z=2$.  The scaling arguments of
Bausch~\etal~(1981) indicate that this is in fact a lower bound on the
value of $z$ for all dimensions up to four, and numerical investigations
have in general measured values slightly higher than this.  The most
accurate determination of $z$ of which we are aware is that of Nightingale
and Bl\"ote~(1996), who measured a value of $z=2.1665\pm0.0012$ for the
dynamic exponent of the two-dimensional Ising model with a local update
algorithm.  Thus the CPU time required to perform a simulation of this
model with a given degree of accuracy increases slightly faster than $L^4$
with system size for temperatures close to the phase transition, which
severely limits the system sizes which can be studied, even with the
fastest computers.

One solution to this problem is to try and find Monte Carlo algorithms
which can update volumes on the order of $\xi^d$ in one step, rather than
by the slow diffusion of domain walls.  Since $\xi$ can become arbitrarily
large, this implies an algorithm with a true non-local update move---one
capable of updating an arbitrary number of spins in one time-step.  In the
first part of this chapter we will study one of the most widely used and
successful such algorithms, the Wolff algorithm, which has a measured
critical exponent close to zero for the Ising model, a great improvement
over the Metropolis case.

\section{The Wolff algorithm}
\label{wolff}
\begin{figure}
\normalfigure{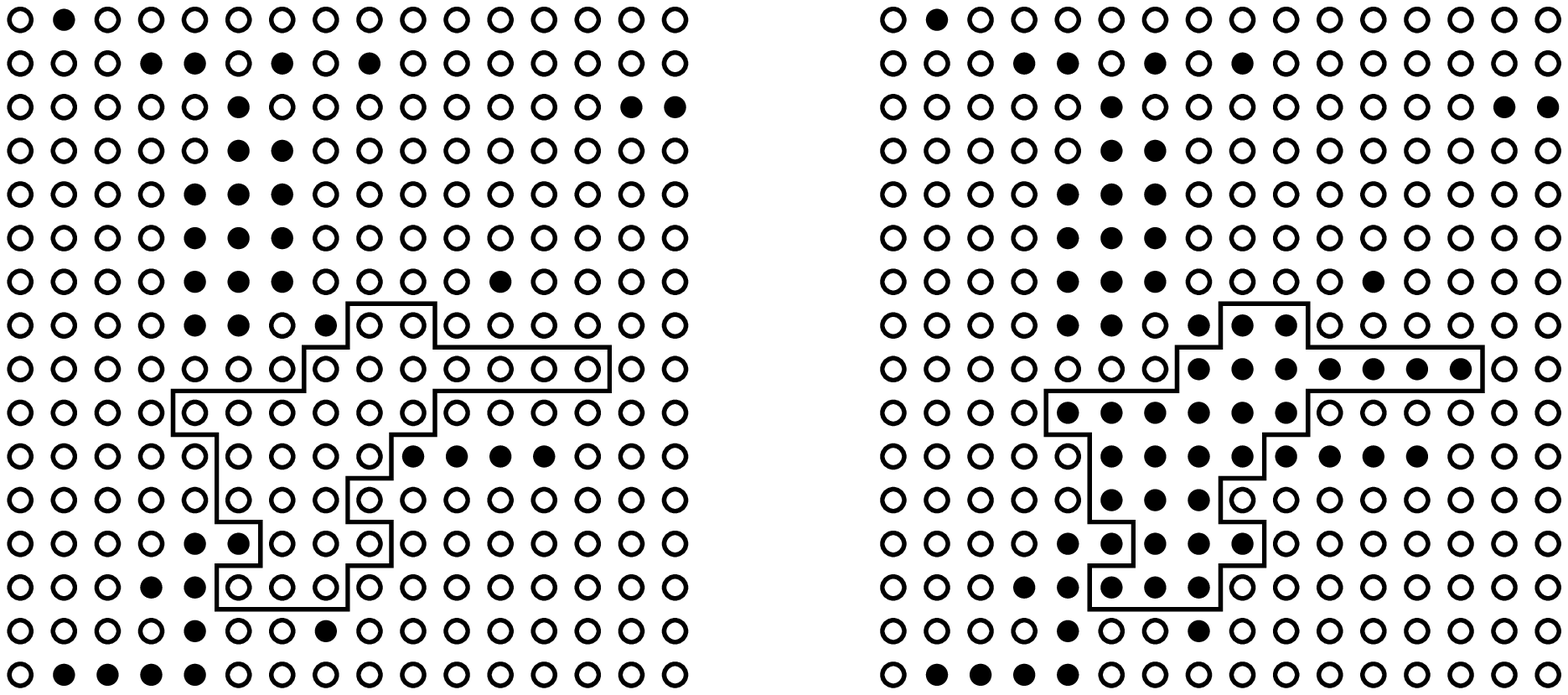}
\capt{Flipping a cluster in a simulation of the 2D Ising model.  The solid
and open circles represent up and down spins in the model.}
\label{frames}
\end{figure}

An elegant non-local Monte Carlo algorithm applicable to the Ising model
has been proposed by Wolff~(1989), based on previous work by Swendsen and
Wang~(1987).  At each Monte Carlo step, this algorithm flips one contiguous
cluster of similarly-oriented spins.  Clusters are built up by starting
from a randomly-chosen seed spin and adding further spins if they are
(a)~adjacent to a spin which already belongs to the cluster and
(b)~pointing in the same direction as the spins in the cluster.  Spins
which satisfy these criteria are added to the cluster with some probability
$P_{\rm add}$.  Eventually the cluster will stop growing when all possible
candidates for addition have been considered and at this point the cluster
is flipped over with some acceptance probability $A$.  It is
straightforward to calculate what value $A$ should take for a given choice
of $P_{\rm add}$.  Consider two states of the system, $\mu$ and $\nu$,
illustrated in Figure~\fref{frames}.  They differ from one another by the
flipping of a single cluster of similarly-oriented spins.  The crucial
thing to notice is the way the spins are oriented around the edge of the
cluster (which is indicated by the line in the figure).  Notice that in
each of the two states, some of the spins just outside the cluster are
pointing the same way as the spins in the cluster.  The bonds between these
spins and the ones in the cluster have to be broken when the cluster is
flipped to reach the other state.  Inevitably, those bonds which are not
broken in going from $\mu$ to $\nu$ must be broken if we make the reverse
move from $\nu$ to $\mu$.

Consider now a cluster Monte Carlo move which takes us from $\mu$ to $\nu$.
There are in fact many such moves---we could choose any of the spins in the
cluster as our seed spin, and then we could add the rest of the spins to it
in a variety of orders.  For the moment, however, let us just consider one
particular move, starting with a particular seed spin and then adding
others to it in a particular order.  Consider also the reverse move, which
takes us back to $\mu$ from $\nu$, starting with the same seed spin, and
adding others to it in exactly the same order as the forward move.  The
probability of choosing the seed is the same in the two directions, as is
the probability of adding each spin to the cluster.  The only thing which
changes between the two is the probability of ``breaking'' bonds around the
edge of the cluster because the bonds which have to be broken are different
in the two cases.  Suppose that, for the forward move, there are $m$ bonds
which have to be broken in order to flip the cluster.  These broken bonds
represent correctly-oriented spins which were not added to the cluster by
the algorithm.  The probability of not adding such a spin is $1-P_{\rm
  add}$.  Thus the probability of not adding all of them, which is
proportional to the selection probability $g(\mu\to\nu)$ for the forward
move, is $(1 - P_{\rm add})^m$.  If there are $n$ bonds which need to be
broken in the reverse move then the probability of doing it will be $(1 -
P_{\rm add})^n$, which is proportional to $g(\nu\to\mu)$.  The condition of
detailed balance then tells us that
\begin{equation}
{g(\mu\to\nu) A(\mu\to\nu)\over g(\nu\to\mu) A(\nu\to\mu)} =
(1 - P_{\rm add})^{m-n} {A(\mu\to\nu)\over A(\nu\to\mu)} =
\e^{-\beta(E_\nu - E_\mu)},
\label{wolffbal}
\end{equation}
where $A(\mu\to\nu)$ and $A(\nu\to\mu)$ are the acceptance ratios for the
moves in the two directions.  The change in energy $E_\nu - E_\mu$ between
the two states also depends on the bonds which are broken.  For each of the
$m$ bonds which are broken in going from $\mu$ to $\nu$, the energy changes
by $+2J$, where $J$ is the interaction strength between neighbouring spins.
For each of the $n$ bonds which are made, the energy changes by $-2J$.
Thus
\begin{equation}
E_\nu - E_\mu = 2J(m-n).
\end{equation}
Substituting into Equation~\eref{wolffbal} and rearranging we derive the
following condition on the acceptance ratios:
\begin{equation}
{A(\mu\to\nu)\over A(\nu\to\mu)} = [\e^{2\beta J} (1 - P_{\rm add})]^{n-m}.
\label{wolffaccept}
\end{equation}
But now we notice a delightful fact: if we choose
\begin{equation}
P_{\rm add} = 1 - \e^{-2\beta J}
\label{padd}
\end{equation}
then the right-hand side of Equation~\eref{wolffaccept} is just 1,
independent of any properties of the states $\mu$ and $\nu$, or the
temperature, or anything else at all.  With this choice, we can make the
acceptance ratios for both forward and backward moves unity, which is the
best possible value they could take.  Every move we propose is accepted and
the algorithm still satisfies detailed balance.  The choice~\eref{padd}
then defines the Wolff cluster algorithm for the Ising model, whose precise
statement would go something like this:
\begin{enumerate}
\setlength{\itemsep}{10pt}
\item Choose a seed spin at random from the lattice.
\item Look in turn at each of the neighbours of that spin.  If they are
  pointing in the same direction as the seed spin, add them to the cluster
  with probability $P_{\rm add} = 1 - \e^{-2\beta J}$.
\item For each spin that was added in the last step, examine each of {\em
    its\/} neighbours to find the ones which are pointing in the same
  direction and add each of them to the cluster with the same probability
  $P_{\rm add}$.  (Notice that as the cluster becomes larger we may find
  that some of the neighbours are already members of the cluster, in which
  case obviously you don't have to consider adding them again.  Also, some
  of the spins may have been considered for addition before, as neighbours
  of other spins in the cluster, but rejected.  In this case, they get
  another chance to be added to the cluster on this step.)  This step is
  repeated as many times as necessary until there are no spins left in the
  cluster whose neighbours have not been considered for inclusion in the
  cluster.
\item Flip all the spins in the cluster.
\end{enumerate}

The flipping of these clusters is a much less laborious task than in the
case of the Metropolis algorithm---as we shall see it takes a time
proportional to the size of the cluster to grow it and then turn it
over---so we have every hope that the algorithm will indeed have a lower
dynamic exponent and less critical slowing down.  In Section~\sref{wolffz}
we show that this is indeed the case.  First we look briefly at how the
Wolff algorithm is implemented.

The standard way to implement the Wolff algorithm is a very straightforward
version of the list of steps given above.  First, we choose at random one
of the spins on the lattice to be our seed.  We look at each of its
neighbours, to see if it points in the same direction as the seed spin.  If
it does then with probability $P_{\rm add}$ we add that spin to the cluster
and we store its coordinates on a stack.  When we have exhausted the
neighbours of the spin we are looking at, we pull a spin off the stack, and
we start checking {\em its\/} neighbours one by one.  Any new spins added
to the cluster are again also added to the stack, and we go on pulling
spins off the stack and looking at their neighbours, until the stack is
empty.  This algorithm guarantees that we consider the neighbours of every
spin added to the cluster, as we should.  Furthermore, the amount of time
spent on each spin in the cluster is, on average, the same.  Each one gets
added to the stack, pulled off again at some later stage and has all of its
neighbours considered as candidates for addition to the cluster.  The time
taken to flip over all the spins in the cluster also goes like the number
of spins, so the entire time taken by one Monte Carlo move in the Wolff
algorithm is proportional to the number of spins in the cluster.

\subsection{Properties of the Wolff algorithm}
\label{properties}
In this section we look at how the Wolff algorithm actually performs in
practice.  As we will see, it performs extremely well when we are near the
critical temperature, but is actually a little slower than the Metropolis
algorithm at very high or low temperatures.

\begin{figure}
\fourfigure{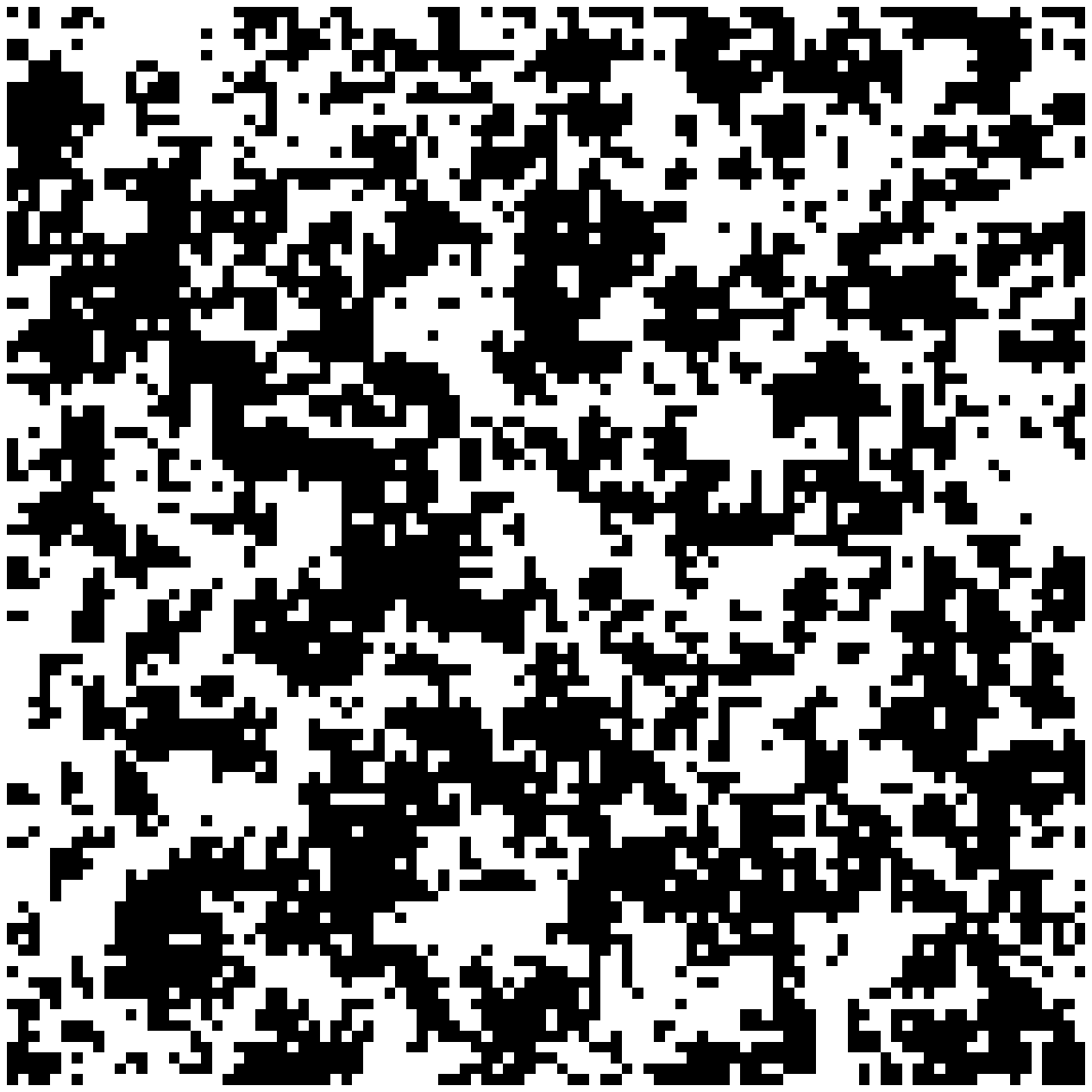}{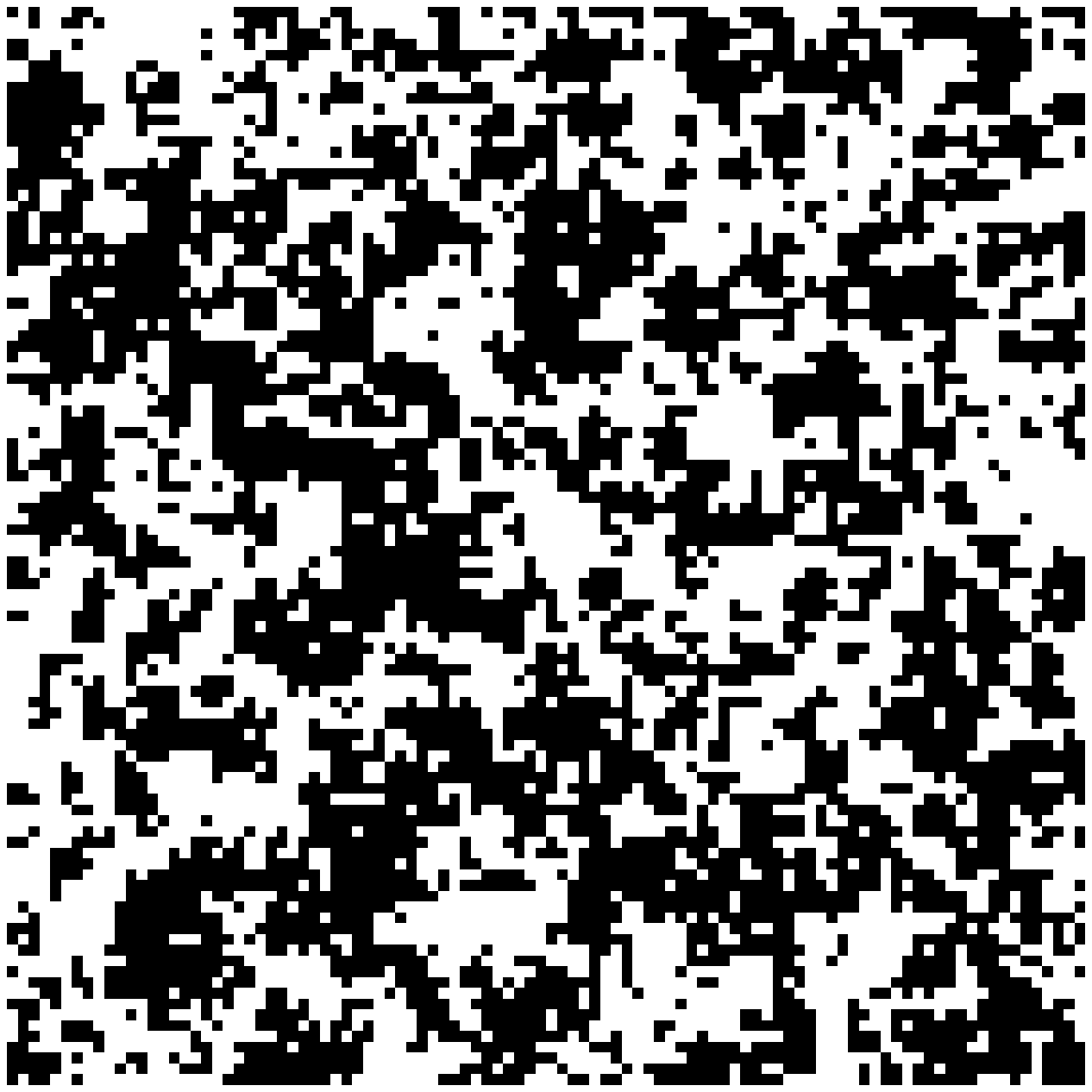}{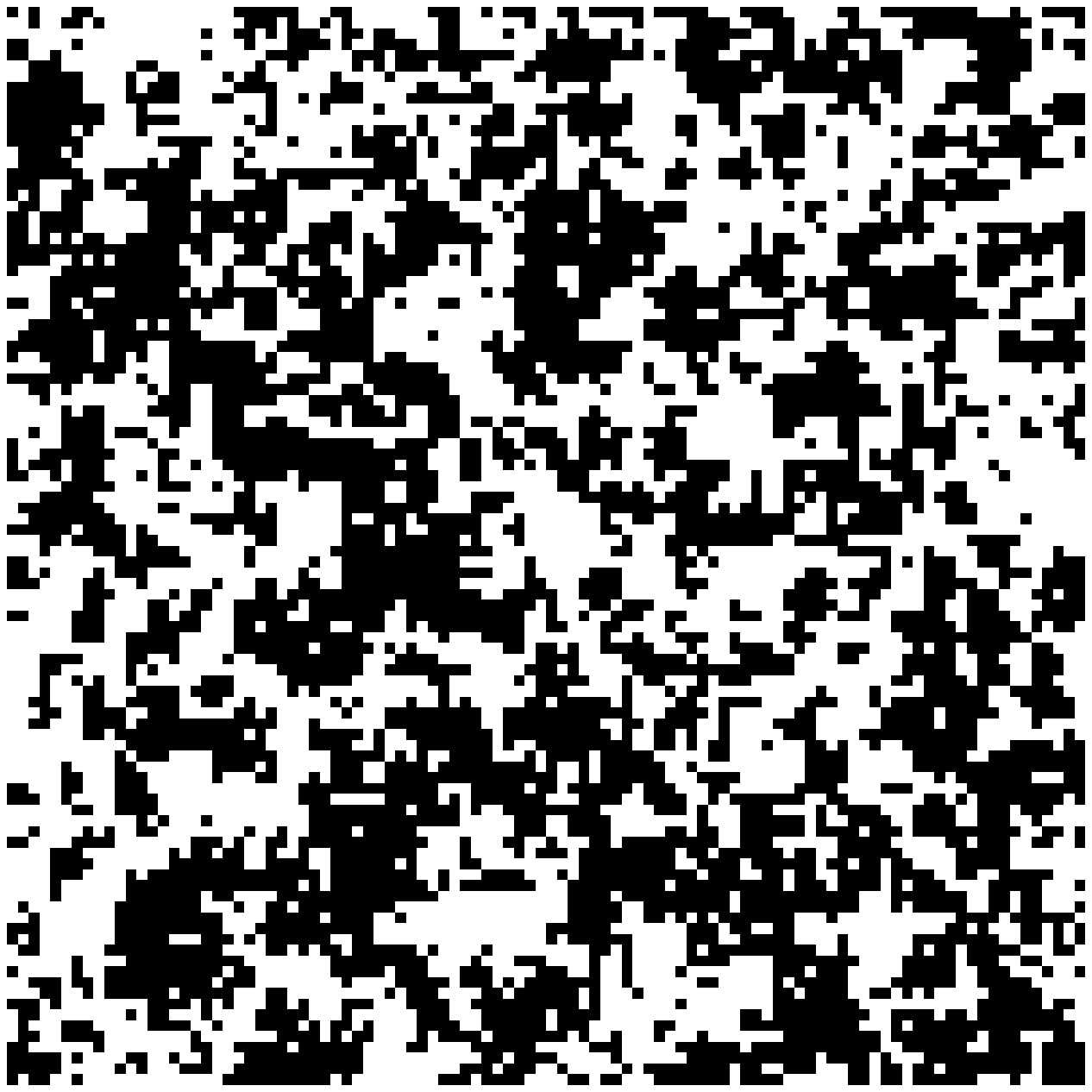}{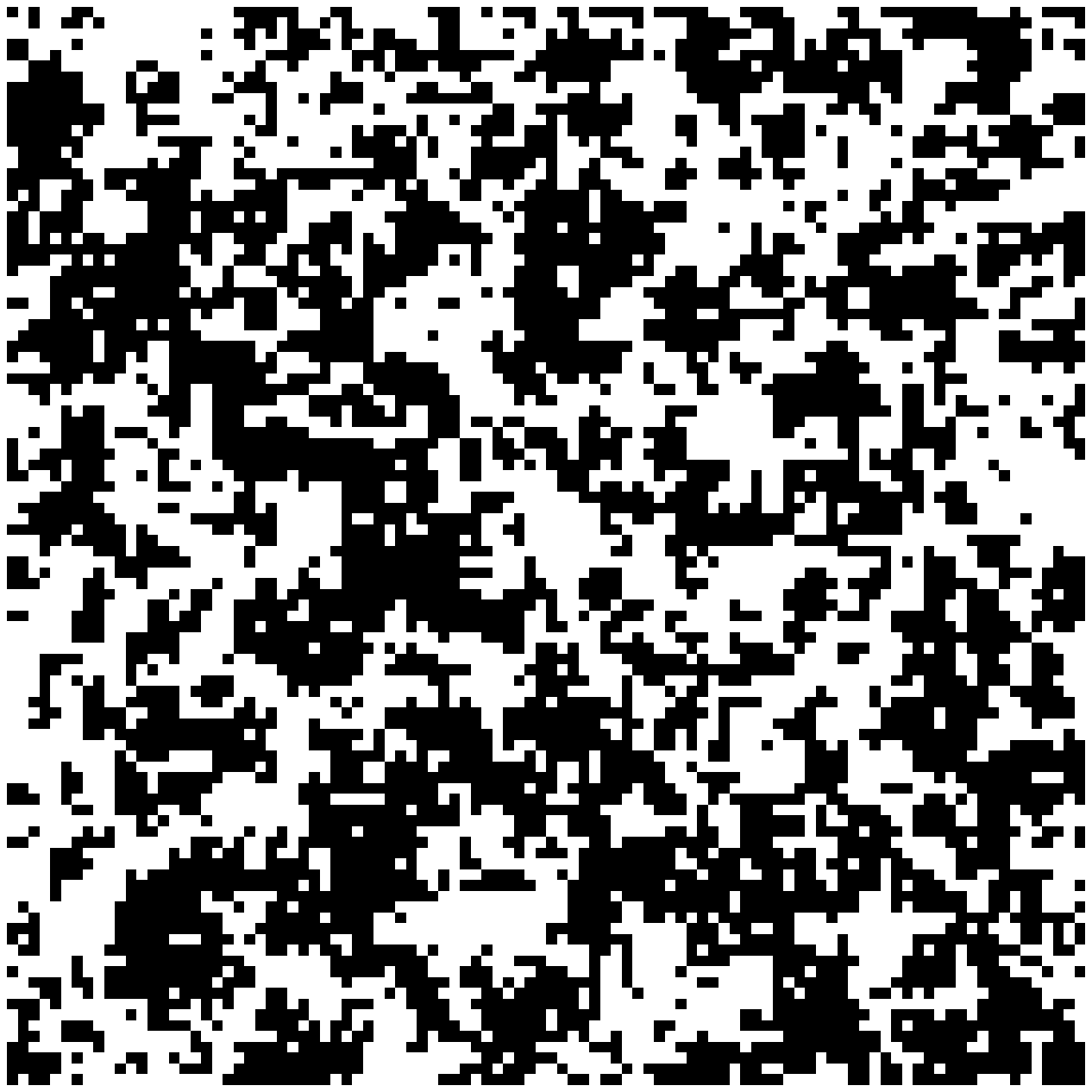}
\bigskip
\fourfigure{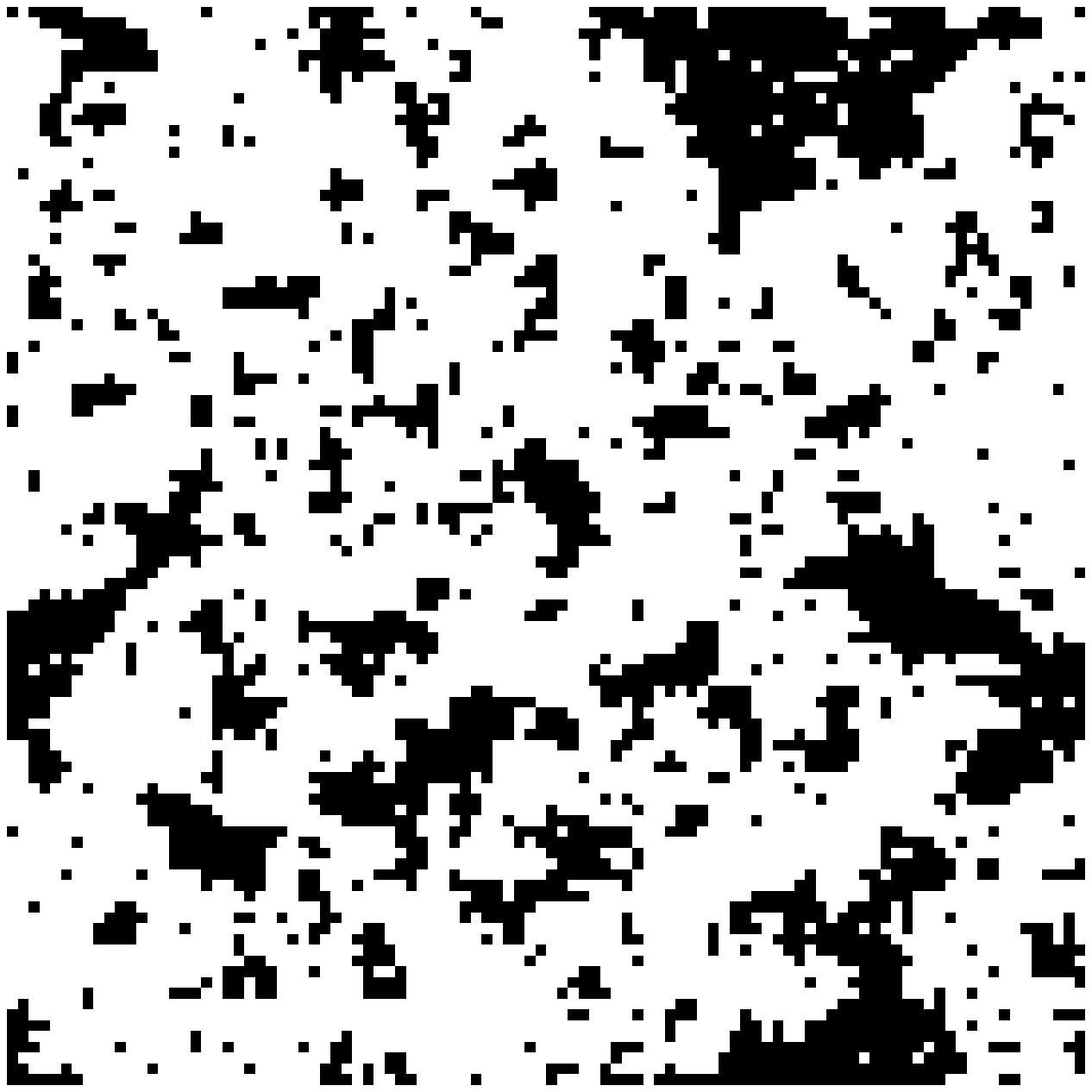}{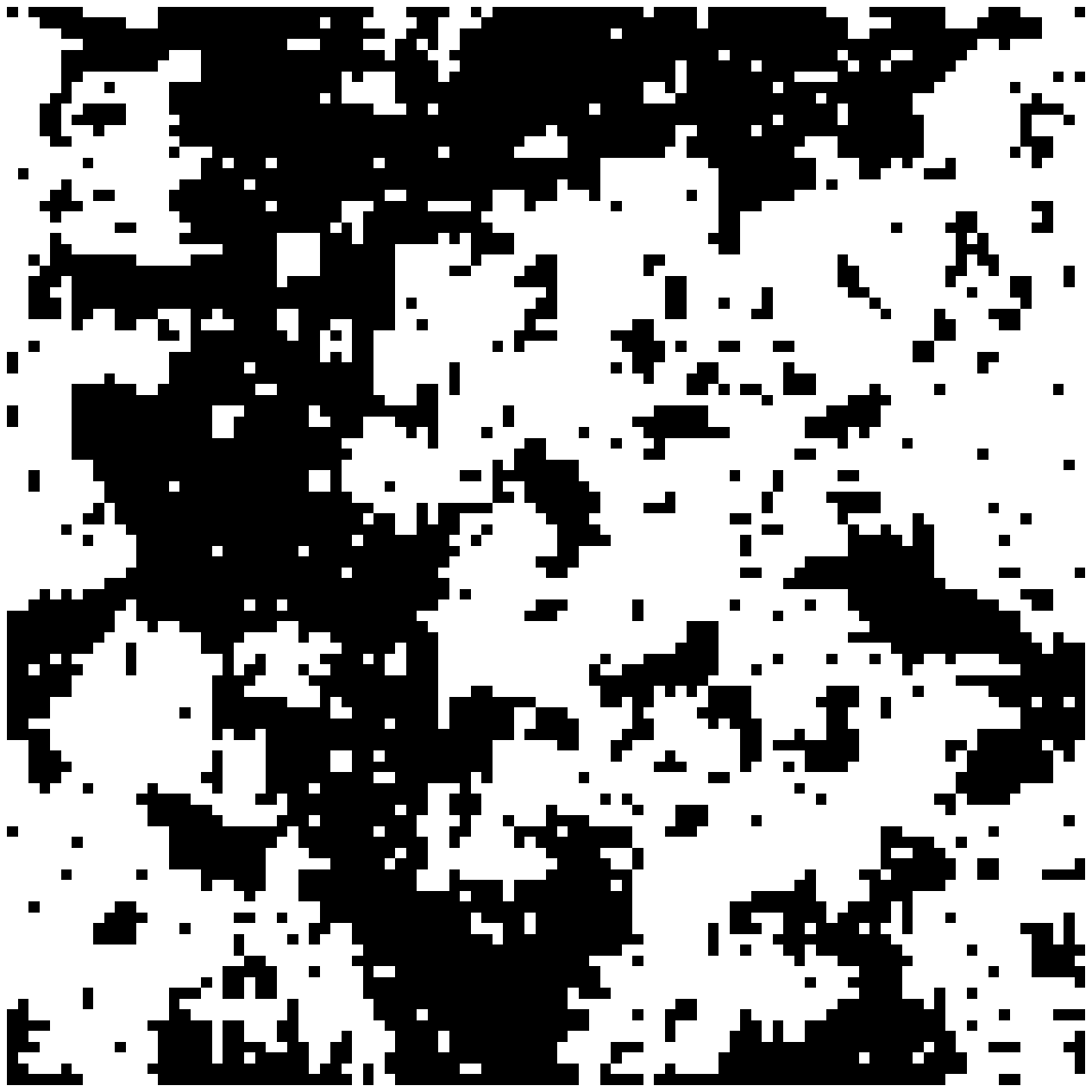}{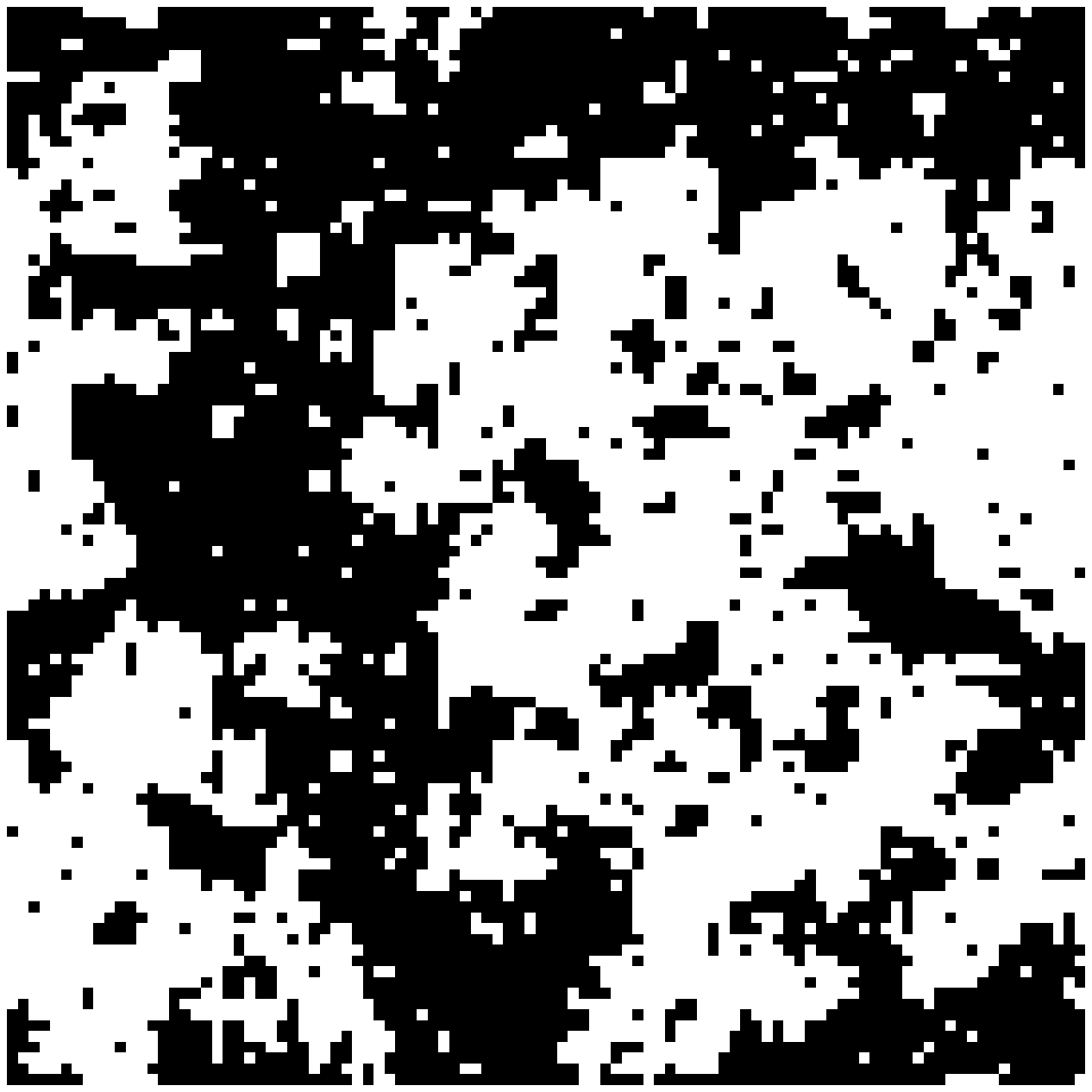}{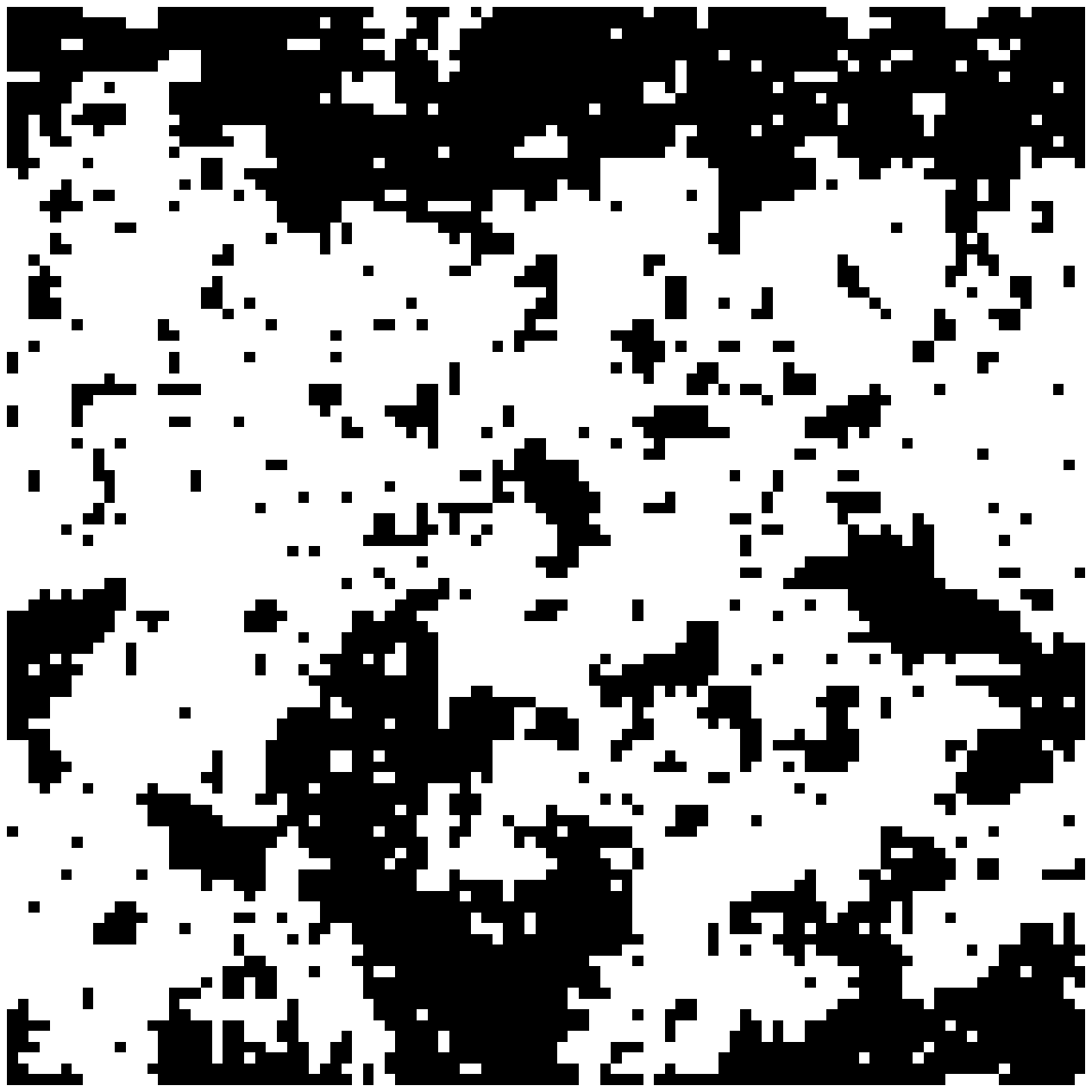}
\bigskip
\fourfigure{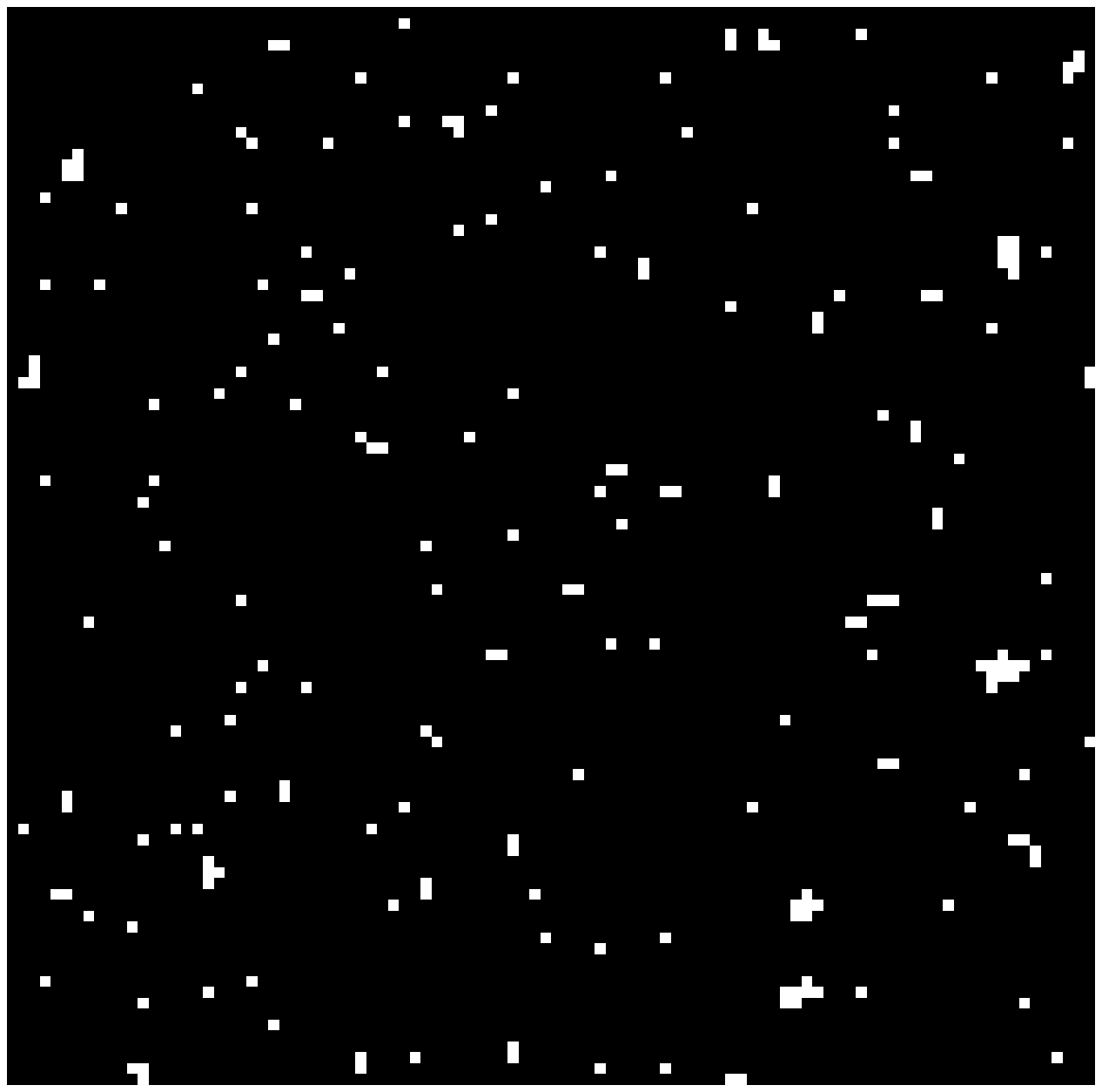}{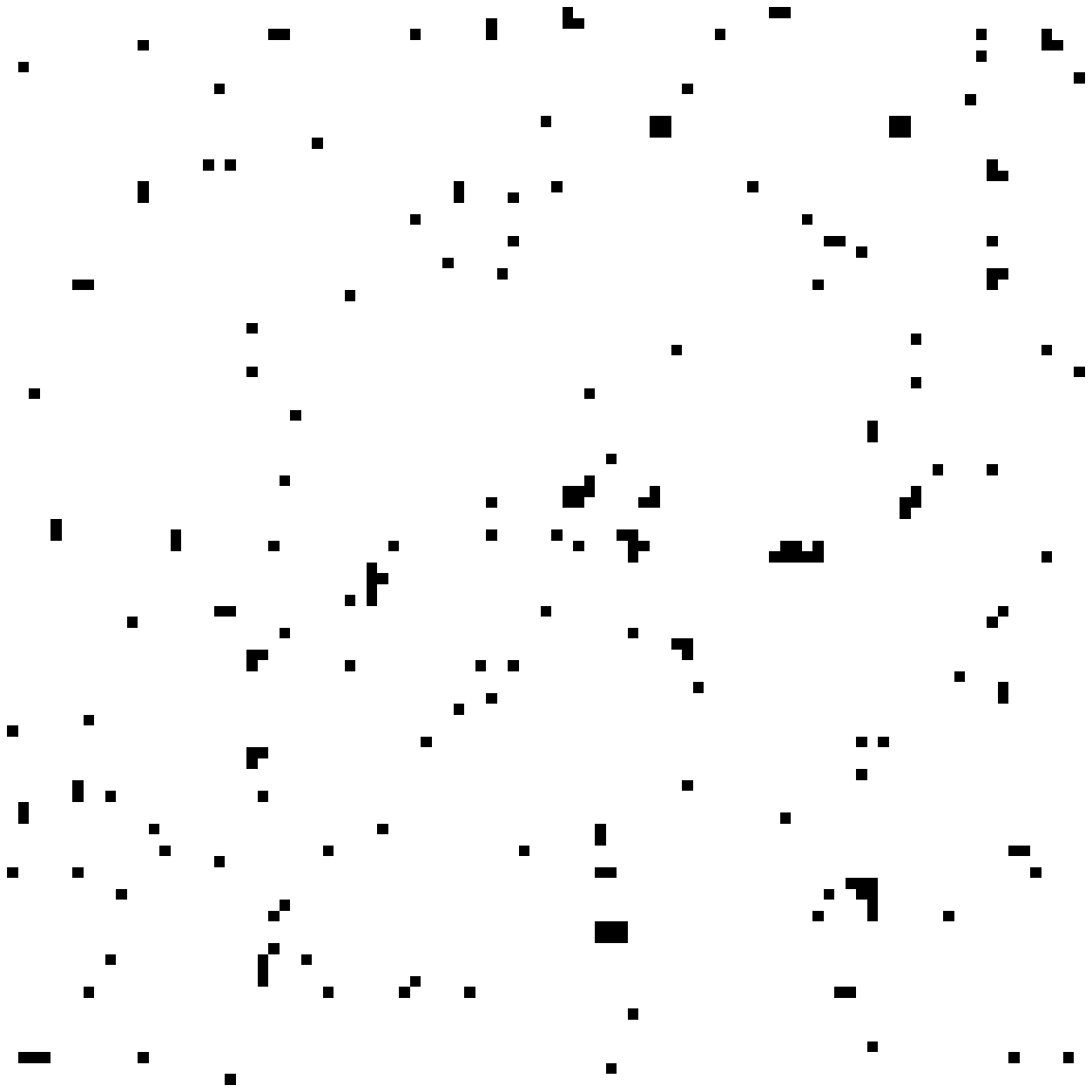}{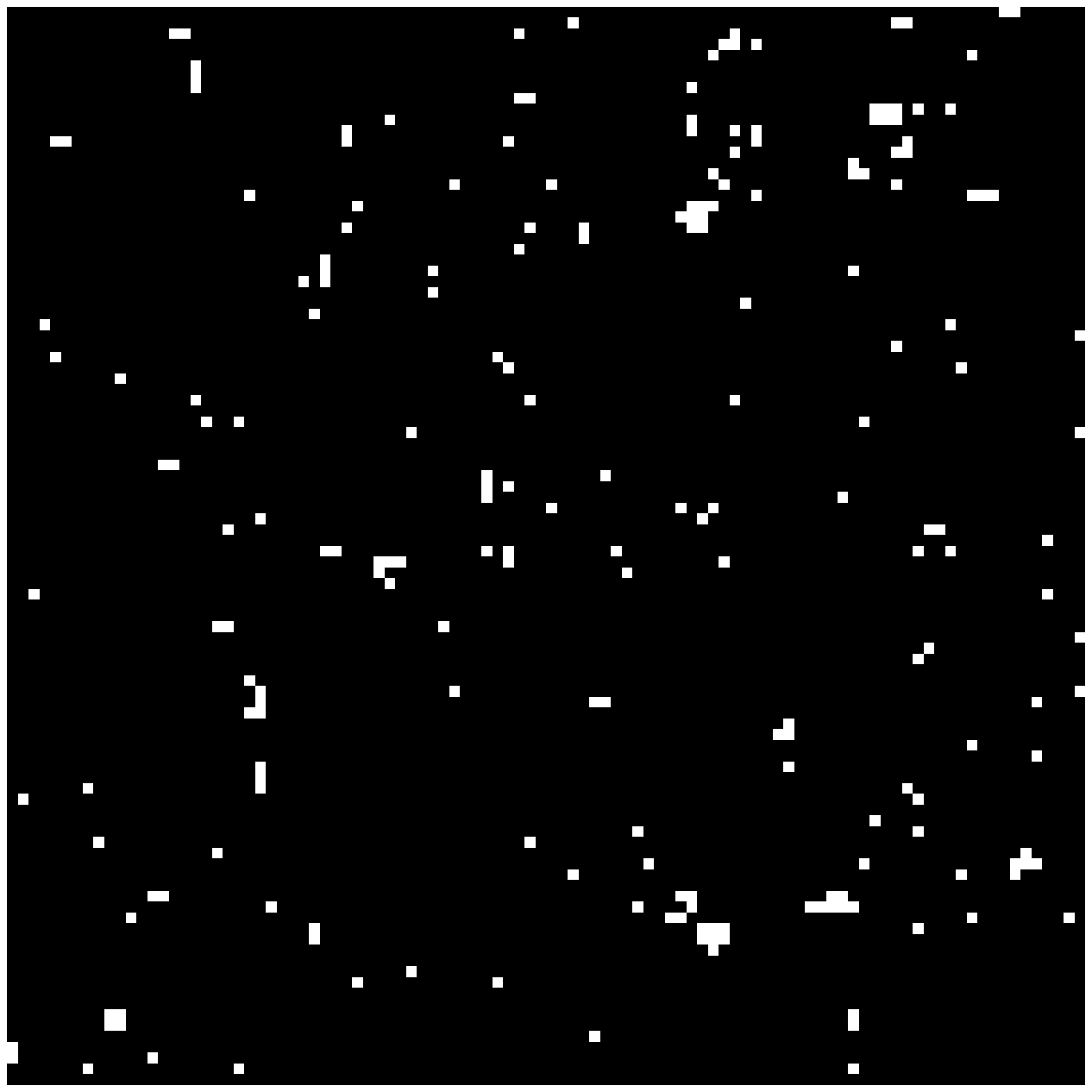}{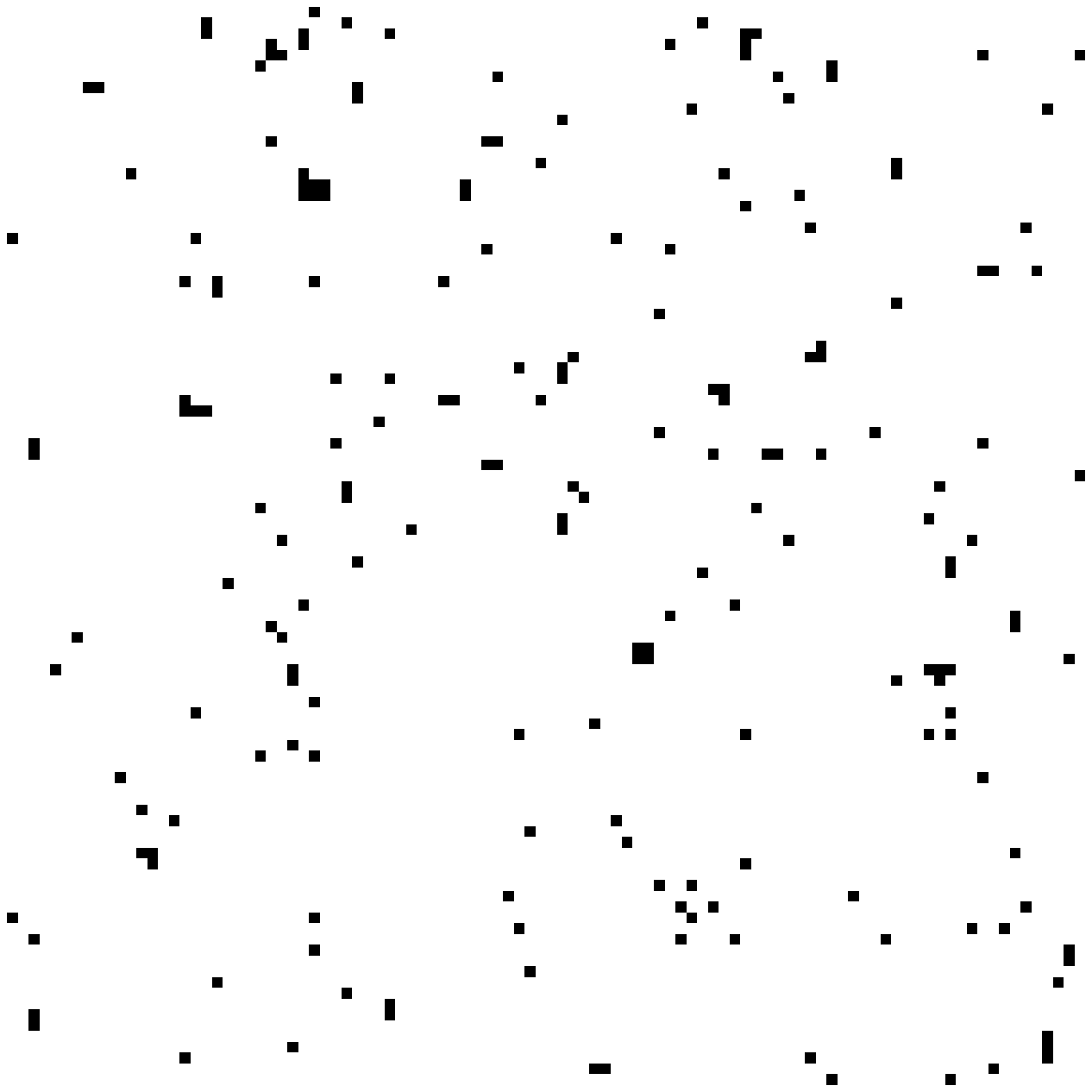}
\capt{Consecutive states of a $100\times100$ Ising model in equilibrium
simulated using the Wolff algorithm.  The top row of four states are at a
temperature $kT=2.8J$, which is well above the critical temperature, the
middle row is close to the critical temperature at $kT=2.3J$, and the
bottom row is well below the critical temperature at $kT=1.8J$.}
\label{wolffsnaps}
\end{figure}

\begin{figure}
\smallfigure{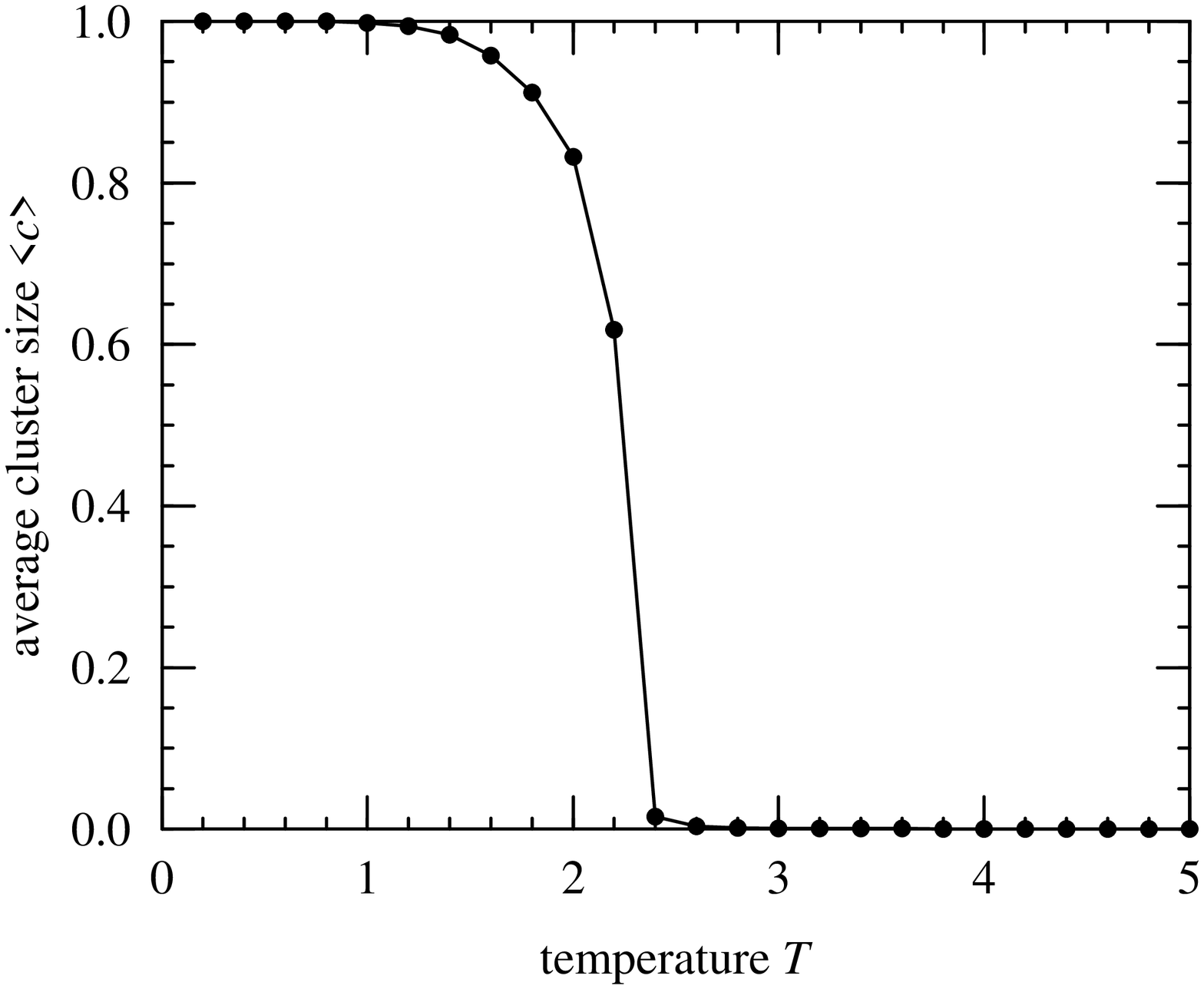}
\capt{Mean cluster size in the Wolff algorithm as a fraction of the size
of the lattice, measured as function of temperature.  The error bars on the
measurements are not shown, because they are smaller than the points.  The
lines are just a guide to the eye.}
\label{csize}
\end{figure}

Figure~\fref{wolffsnaps} shows a series of states generated by the
algorithm at each of three temperatures, one above $T_c$, one close to
$T_c$, and one below it.  Up and down spins are represented by the black
and white dots.  Consider first the middle set of states, the ones near
$T_c$.  If you examine the four frames, it is not difficult to make out
which cluster flipped at each step.  Clearly the algorithm is doing its
job, flipping large areas of spins when we are in the critical region.  In
the $T>T_c$ case, it is much harder to make out the changes between one
frame and the next.  The reason for this is that in this temperature region
$P_{\rm add}$ is quite small (see Equation~\eref{padd}) and this in turn
makes the clusters small, so it is hard to see when they flip over.  Of
course, this is exactly what the algorithm is supposed to do, since the
correlation length here is small, and we don't expect to get large regions
of spins flipping over together.  In Figure~\fref{csize}, we have plotted
the mean size of the clusters flipped by the Wolff algorithm over a range
of temperatures, and, as we can see, it does indeed become small at large
temperatures.  When the temperature gets sufficiently high (around $kT=10$
in two dimensions), the mean size of the clusters becomes hardly greater
than a single spin.  In other words, the single seed spin for each cluster
is being flipped over with probability one at each step, but none of its
neighbours are.  However, this is just exactly what the single-spin flip
Metropolis algorithm does in this temperature regime.  When $T$ is large,
the Metropolis acceptance ratio is 1, or very close to it, for any
transition between two states $\mu$ and $\nu$.  Thus in the limit of high
temperatures, the Wolff algorithm and the Metropolis algorithm become the
same thing.  But notice that the Wolff algorithm will actually be the
slower of the two in this case, because for each seed spin it has to go
through the business of testing each of the neighbours for possible
inclusion in the cluster, whereas the Metropolis algorithm only has to
decide whether to flip a single spin or not, a comparatively simple
computational task.  Thus, even if the Wolff algorithm is a good thing near
to the phase transition (which it is), there comes a point as the
temperature increases where the Metropolis algorithm becomes better.

Now let us turn to the simulation at low $T$.  Looking at the bottom row of
Figure~\fref{wolffsnaps}, the action of the algorithm is dramatically
obvious in this case---almost every spin on the lattice is being flipped at
each step.  The reason for this is clear.  When we are well below the
critical temperature the Ising model develops a backbone of
similarly-oriented spins which spans the entire lattice.  When we choose
our seed spin for the Wolff algorithm, it is likely that we will land on
one of the spins comprising this backbone.  Furthermore, the probability
$P_{\rm add}$, Equation~\eref{padd}, is large when $T$ is small, so the
neighbours of the seed spin are not only likely to be aligned with it, but
are also likely to be added to the growing cluster.  The result is that the
cluster grows (usually) to fill almost the entire backbone of spontaneously
magnetized spins, and then they are all flipped over in one step.

On the face of it, this seems like a very inefficient way to generate
states for the Ising model.  After all, we know that what should really be
happening in the Ising model at low temperature is that most of the spins
should be lined up with one another, except for a few excitation spins,
which are pointing the other way (see the Figure~\fref{wolffsnaps}).  Every
so often, one of these excitation spins flips back over to join the
majority pointing the other way, or perhaps one of the backbone spins gets
flipped by a particularly enthusiastic thermal excitation and becomes a new
excitation spin.  This of course is exactly what the Metropolis algorithm
does in this regime.  The Wolff algorithm on the other hand is removing the
single-spin excitations by the seemingly extravagant measure of {\em
  flipping all the other spins in the entire lattice\/} to point the same
way as the single lonely excitation.  In fact, however, it's actually not
such a stupid thing to do.  First, let us point out that, since the Wolff
algorithm never flips spins which are pointing in the opposite direction to
the seed spin, all the excitation spins on the entire lattice end up
pointing the same way as the backbone when the algorithm flips over a
percolating cluster.  Therefore, the Wolff algorithm gets rid of {\em
  all\/} the excitation spins on the lattice in a single step.  Second,
since we know that the Wolff algorithm generates states with the correct
Boltzmann probabilities, it must presumably create some new excitation
spins at the same time as it gets rid of the old ones.  And indeed it does
do this, as is clear from the frames in the figure.  Each spin in the
backbone gets a number of different chances to be added to the cluster; for
most of the spins this number is just the lattice coordination number $z$,
which is four in the case of the square lattice.  Thus the chance that a
spin will not be added to the cluster at all is $(1-P_{\rm add})^4$.  This
is a small number, but still non-zero.  (It is $0.012$ at the temperature
$kT=1.8J$ used in the figure.)  Thus there will be a small number of spins
in the backbone which get left out of the percolating cluster and are not
flipped along with the others.  These spins become the new excitations on
the lattice.

The net result of all this is that after only one Wolff Monte Carlo step,
all the old excitations have vanished and a new set have appeared.  This
behaviour is clear in Figure~\fref{wolffsnaps}.  At low temperatures, the
Wolff algorithm generates a complete new configuration of the model at
every Monte Carlo step, though the payoff is that it has to flip very
nearly every spin on every step too.  In the Metropolis algorithm at low
temperatures it turns out that you have to do about one Monte Carlo step
per site (each one possibly flipping one spin) to generate a new
independent configuration of the lattice.  To see this, we need only
consider again the excitation spins.  The average acceptance ratio for
flipping these over will be close to 1 in the Metropolis algorithm, because
the energy of the system is usually lowered by flipping them.  (Only on the
rare occasions when several of them happen to be close together might this
not be true.)  Thus, it will on average just take $N$ steps, where $N$ is
the number of sites on the lattice, before we find any particular such spin
and flip it---in other words, one Monte Carlo step per site.  But we know
that the algorithm correctly generates states with their Boltzmann
probability, so, in the same $N$ steps that it takes to find all the
excitation spins and flip them over to join the backbone, the algorithm
must also choose a roughly equal number of new spins to excite out of the
backbone, just as the Wolff algorithm also does.  Thus, it takes one sweep
of the lattice to replace all the excitation spins with a new set,
generating an independent state of the lattice.  (This is in fact the best
possible performance that any Monte Carlo algorithm can have, since one has
to allow at least one sweep of the lattice in order to generate a new
configuration, so that each spin gets the chance to change its value.)

Again then, we find that the Wolff algorithm and the Metropolis algorithm
are roughly comparable, this time at low temperatures.  Both need to
consider about $N$ spins for flipping in order to generate an independent
state of the lattice.  Again, however, the additional complexity of the
Wolff algorithm is its downfall, and the extremely simple Metropolis
algorithm has the edge in speed.

So, if the Metropolis algorithm beats the Wolff algorithm (albeit only by
a slim margin) at both high and low temperatures, that leaves only the
intermediate regime, close to $T_c$ in which the Wolff algorithm might be
worthwhile.  This of course is the regime in which we designed the
algorithm to work well, so we have every hope that it will beat the
Metropolis algorithm there, and indeed it does, very handily.

\subsection{Correlation time}
\label{wolffz}
Before we measure the correlation time of the Wolff algorithm, we need to
consider exactly how it should be defined.  If we are going to compare the
correlation times of the Wolff algorithm and the Metropolis algorithm near
to the phase transition as a way of deciding which is the better algorithm
in this region, it clearly would not be fair to measure it for both
algorithms in terms of number of Monte Carlo steps.  A single Monte Carlo
step in the Wolff algorithm is a very complicated procedure, flipping maybe
hundreds of spins and potentially taking quite a lot of CPU time, whereas
the Metropolis Monte Carlo step is a very simple, quick thing---we can do a
million of them in a second on a good computer---though each one only flips
at most one spin.

In Section~\sref{wolff} we showed that the time taken to complete one step
of the Wolff algorithm is proportional to the number of spins $c$ in the
cluster.  Such a cluster covers a fraction $c/L^d$ of the entire lattice,
and so on average each Monte Carlo step will take an amount of CPU time
comparable to $\av{c}/L^d$ sweeps of the lattice using the single-spin-flip
Metropolis algorithm.  Thus the correct way to define the correlation time
is to write $\tau \propto \tau_{\rm steps} \av{c}/L^d$, where $\tau_{\rm
  steps}$ is the correlation time measured in steps (i.e.,~clusters
flipped) in the Wolff algorithm.  The conventional choice for the constant
of proportionality is 1.  This makes the correlation times for the Wolff
and Metropolis algorithms equal in the limits of low and high temperature,
for the reasons discussed in the last section.  This is not quite fair,
since, as we pointed out earlier, the Wolff algorithm is a little slower
than the Metropolis in these regimes because of its greater complexity.
However, this difference is slight compared with the enormous difference in
performance between the two algorithms near the phase transition which we
will witness in a moment, so for all practical purposes we can write
\begin{equation}
\tau = \tau_{\rm steps} {\av{c}\over L^d}.
\label{wolfftau}
\end{equation}

In our own simulations of the 2D Ising model on a $100\times100$ square
lattice using the two algorithms, we measure the correlation time at $T_c$
of the Wolff algorithm to be $\tau = 2.80\pm0.03$ spin-flips per site, by
contrast with the Metropolis algorithm which has $\tau = 2570\pm330$.  A
factor of a thousand certainly out-weighs any difference in the relative
complexity of the two algorithms.  It is this impressive performance on the
part of the Wolff algorithm which makes it a worthwhile one to use if we
are interested in the behaviour of the model close to $T_c$.

\begin{figure}
\smallfigure{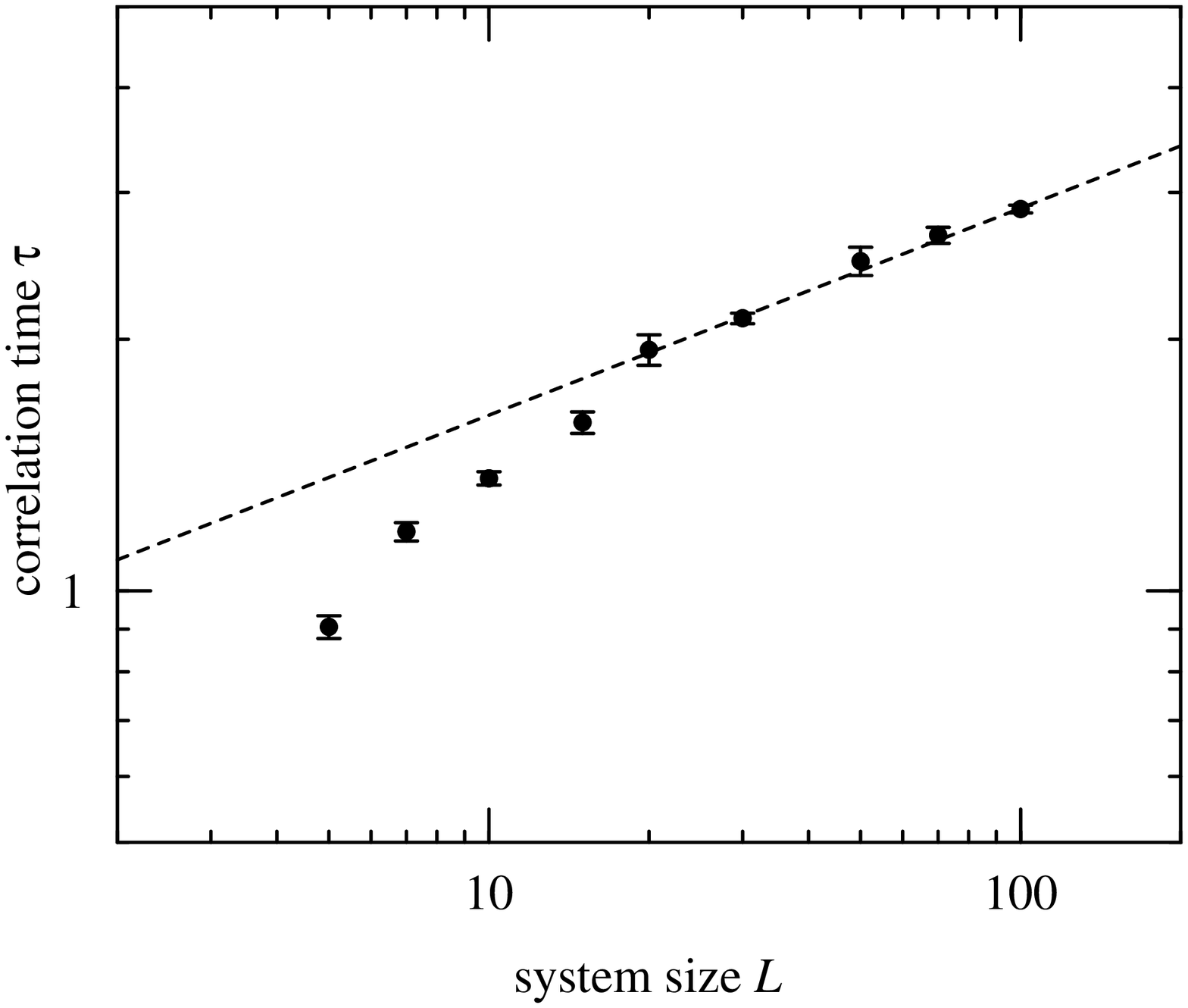}
\capt{The correlation time $\tau$ for the 2D Ising model simulated using
the Wolff algorithm.  The measurements deviate from a straight line
for small system sizes $L$, but a fit to the larger sizes, indicated by
the dashed line, gives a reasonable figure of $z=0.25\pm0.02$ for the
dynamic exponent of the algorithm.}
\label{taulw}
\end{figure}

In Figure~\fref{taulw} we have plotted on logarithmic scales the
correlation time of the Wolff algorithm for the 2D Ising model at the
critical temperature, over a range of different system sizes.  The slope of
the line gives us an estimate of the dynamic exponent.  Our best fit, given
the errors on the data points is $z=0.25\pm0.02$.  This was something of a
rough calculation, though our result is competitive with other more
thorough ones.  The best available figure at the time of writing was that
of Coddington and Baillie~(1992) who measured $z=0.25\pm0.01$.  This figure
is clearly much lower than the $z=2.17$ of the Metropolis algorithm, and
gives us a quantitative measure of how much better the Wolff algorithm
really is.

\subsection{The dynamic exponent and the susceptibility}
\label{zcpu}
In fact in studies of the Wolff algorithm for the 2D Ising model one does
not usually bother to make use of Equation~\eref{wolfftau} to calculate
$\tau$.  If we measure time in Monte Carlo steps (that is, simple cluster
flips), we can define the corresponding dynamic exponent $z_{\rm steps}$ in
terms of the correlation time $\tau_{\rm steps}$ of
Equation~\eref{wolfftau} thus:
\begin{equation}
\tau_{\rm steps} \sim \xi^{-z_{\rm steps}}.
\label{eqtausteps}
\end{equation}
The exponent $z_{\rm steps}$ is related to the real dynamic exponent $z$
for the algorithm by
\begin{equation}
z = z_{\rm steps} + {\gamma\over\nu} - d,
\label{eqwolffz}
\end{equation}
where $\gamma$ and $\nu$ are the critical exponents governing the
divergences of the magnetic susceptibility and the correlation length.  If
we know the values of $\nu$ and $\gamma$, as we do in the case of the 2D
Ising model, then we can use Equation~\eref{eqwolffz} to calculate $z$
without having to measure the mean cluster size in the algorithm, which
eliminates one source of error in the measurement.\footnote{In cases, such
  as the 3D Ising model, for which we don't know the exact values of the
  critical exponents, it may be better to make use of
  Equation~\eref{wolfftau} and measure $z$ directly, as we did in
  Figure~\fref{taulw}.}

The first step in demonstrating Equation~\eref{eqwolffz} is to prove
another useful result, about the magnetic susceptibility $\chi$.  It turns
out that, for temperatures $T\ge T_c$, the susceptibility is related to the
mean size $\av{c}$ of the clusters flipped by the Wolff algorithm thus:
\begin{equation}
\chi = \beta \av{c}.
\label{wolffchi1}
\end{equation}
In many simulations using the Wolff algorithm, the susceptibility is
measured using the mean cluster size in this way.

\begin{figure}
\sidefigure{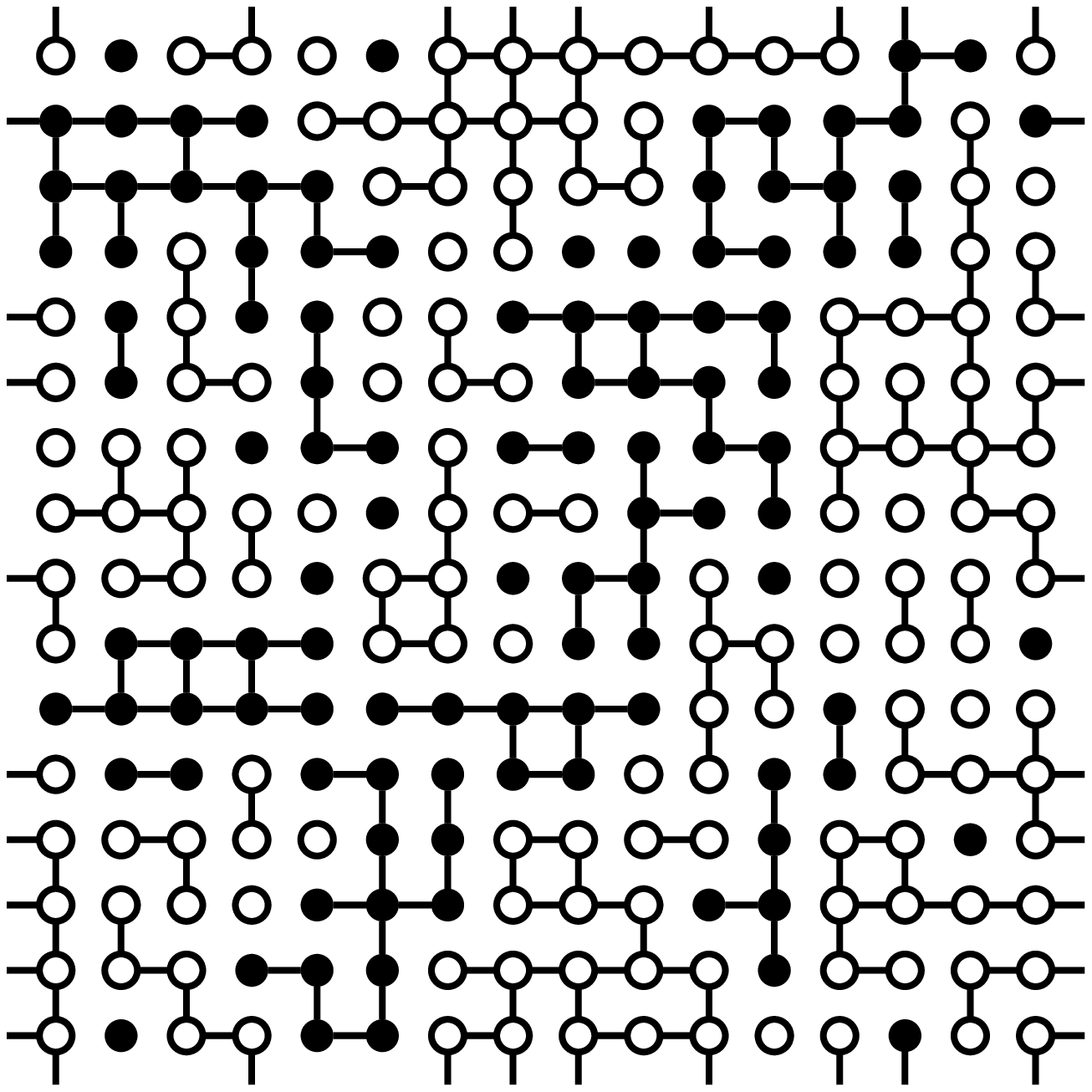}{The whole lattice can be divided into clusters simply
by putting ``links'' down with probability $P_{\rm add}$ between any two
neighbouring spins which are pointing in the same direction.
\label{clusters}}
\end{figure}

The demonstration of Equation~\eref{wolffchi1} goes like this.  Instead of
implementing the Wolff algorithm in the way described in this chapter,
imagine instead doing it a slightly different way.  Imagine that at each
step we look at the whole lattice and for every pair of neighbouring spins
which are pointing in the same direction, we make a ``link'' between them
with probability $P_{\rm add} = 1 - \e^{-2\beta J}$.  When we are done, we
will have divided the whole lattice into many different clusters of spins,
as shown in Figure~\fref{clusters}, each of which will be a correct Wolff
cluster (since we have used the correct Wolff probability $P_{\rm add}$ to
make the links).  Now we choose a single seed spin from the lattice at
random, and flip the cluster to which it belongs.  Then we throw away all
the links we have made and start again.  The only difference between this
algorithm and the Wolff algorithm as we described it is that here we make
the clusters first and choose the seed spin afterwards, rather than the
other way around.  This would not be a very efficient way of implementing
the Wolff algorithm, since it requires us to create clusters all over the
lattice, almost all of which never get flipped, but it is a useful device
for proving Equation~\eref{wolffchi1}.  Why?  Well, we can write the total
magnetization $M$ of the lattice as a sum over all the clusters on the
lattice thus:
\begin{equation}
M = \sum_i S_i c_i.
\label{swm}
\end{equation}
Here $i$ labels the different clusters, $c_i$ is their size (a positive
integer), and $S_i = \pm1$ depending on whether the \th{i} cluster is
pointing up or down, thereby making either a positive or a negative
contribution to $M$.  The mean square magnetization is then
\begin{equation}
\av{M^2} = \Bigl\langle\sum_i S_i c_i \sum_j S_j c_j\Bigr\rangle
         = \Bigl\langle\sum_{i\ne j} S_i S_j c_i c_j\Bigr \rangle +
           \Bigl\langle\sum_i S_i^2 c_i^2\Bigr\rangle.
\end{equation}
Now, since the average of the magnetization is zero under the Wolff
algorithm (this is even true below $T_c$, since the algorithm flips most of
the spins on the lattice at every step at low temperature), the first term
in this expression is an average over a large number of quantities which
are randomly either positive or negative and which will therefore tend to
average out to zero.  The second term on the other hand is an average over
only positive quantities and therefore is not zero.\footnote{Strictly, the
  second term scales like the number of spin configurations in the average
  and the first scales like its square root, so the second term dominates
  for a large number of configurations.} Noting that $S_i^2 = 1$ for all
$i$, we can then write
\begin{equation}
\av{M^2} = \Bigl\langle\sum_i c_i^2\Bigr\rangle,
\end{equation}
or alternatively, in terms of the magnetization per spin $m$:
\begin{equation}
\av{m^2} = {1\over N^2} \Bigl\langle\sum_i c_i^2\Bigr\rangle.
\label{swmsq}
\end{equation}

Now consider the average $\av{c}$ of the size of the clusters which get
flipped in the Wolff algorithm.  This is not quite the same thing as the
average size $\av{c_i}$ of the clusters over the entire lattice, because
when we choose our seed spin for the Wolff algorithm it is chosen at random
from the entire lattice, which means that the probability $p_i$ of it
falling in a particular cluster $i$, is proportional to the size of that
cluster:
\begin{equation}
p_i = {c_i\over N}.
\end{equation}
The average cluster size in the Wolff algorithm is then given by the
average over the probability of the cluster being chosen times the size of
that cluster:
\begin{equation}
\av{c} = \Bigl\langle\sum_i p_i c_i\Bigr\rangle = {1\over N} \Bigl\langle
\sum_i c_i^2\Bigr\rangle = N\av{m^2}.
\end{equation}
Given that $\av{m}=0$ for $T\ge T_c$, we now have our result for the
magnetic susceptibility as promised:
\begin{equation}
\chi = \beta \av{c}.
\label{wolffchi2}
\end{equation}

We can now use this equation to rewrite Equation~\eref{wolfftau} thus:
\begin{equation}
\tau = \tau_{\rm steps} {\chi\over\beta L^d},
\end{equation}
which implies in turn that
\begin{equation}
\xi^{-z} \sim \xi^{-z_{\rm steps}} \xi^{-\gamma/\nu} L^{-d}.
\end{equation}
Near the critical point, this implies that
\begin{equation}
L^z \sim L^{z_{\rm steps}} L^{\gamma/\nu} L^{-d}
\end{equation}
and thus
\begin{equation}
z = z_{\rm steps} + {\gamma\over\nu} - d
\label{wolffz2}
\end{equation}
as we suggested earlier on.

\begin{figure}
\smallfigure{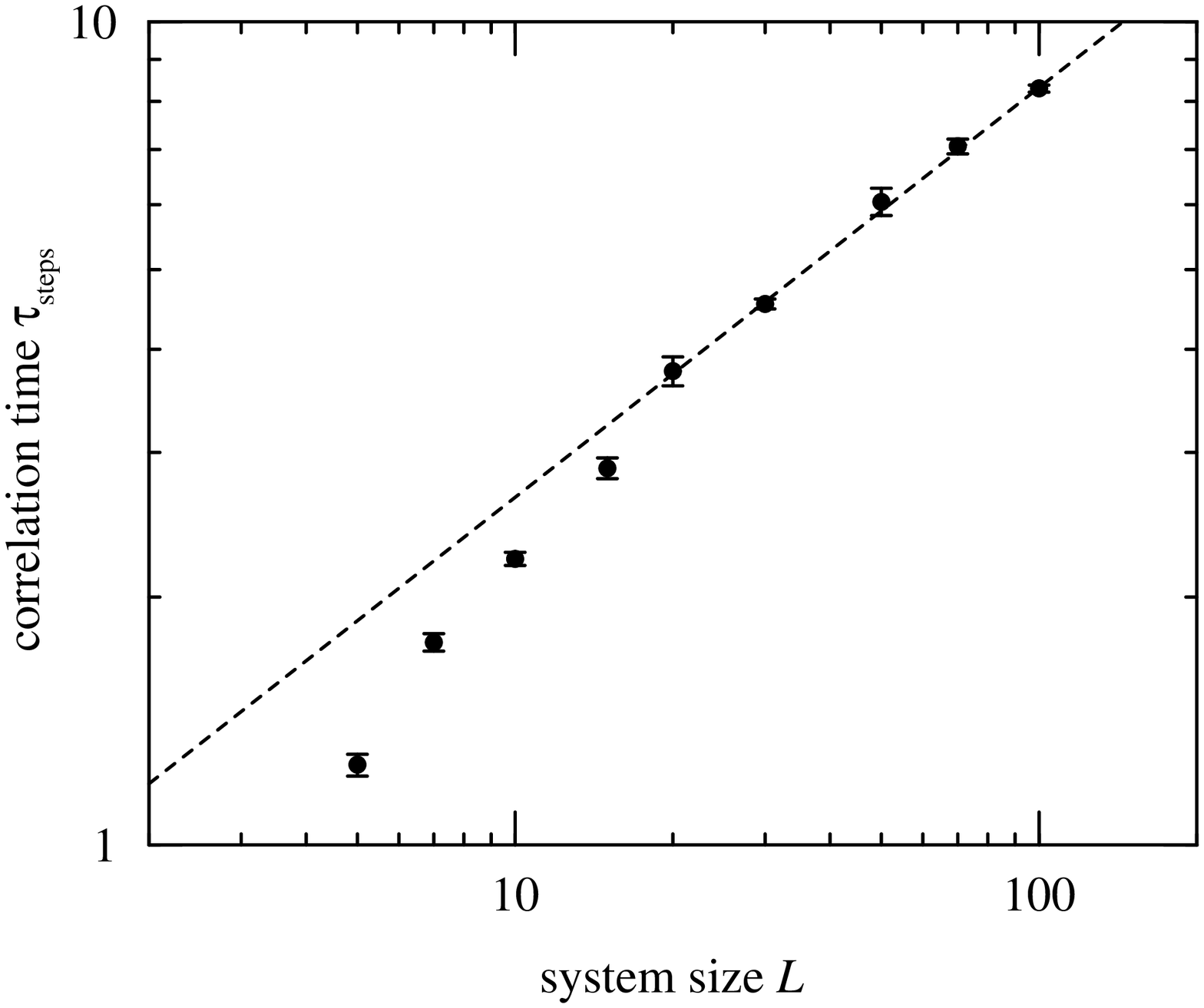}
\capt{The correlation time $\tau_{\rm steps}$ of the 2D Ising model
simulated using the Wolff algorithm, measured in units of Monte Carlo steps
(i.e.,~cluster flips).  The fit gives us a value of $z_{\rm steps} =
0.50\pm0.01$ for the corresponding dynamic exponent.}
\label{tausteps}
\end{figure}

In Figure~\fref{tausteps} we have replotted the results from our Wolff
algorithm simulations of the 2D Ising model using the correlation time
measured in Monte Carlo steps (i.e.,~cluster flips).  The best fit to the
data gives a figure of $z_{\rm steps} = 0.50\pm0.01$.  Using the accepted
values $\nu=1$ and $\gamma=\frac74$ for the 2D Ising model (Onsager~1944),
and setting $d=2$, Equation~\eref{wolffz2} then gives us
$z=0.25\pm0.01$, which is as good as the best published result for this
exponent.

\section{Further algorithms for the Ising model}
\label{further}
We have looked in detail at the Wolff algorithm for simulating the Ising
model.  Close to $T_c$ this algorithm is significantly more efficient than
the Metropolis one; although it is more complex than the Metropolis
algorithm, the Wolff algorithm has a very small dynamic exponent, which
means that the time taken to perform a simulation scales roughly like the
size of the system, which is the best that we can hope for in any
algorithm.

However, many other algorithms have been suggested for the simulation of
the Ising model.  In this section we talk about a few of these alternative
algorithms.

\subsection{The Swendsen-Wang algorithm}
\label{swalg}
After the Metropolis and Wolff algorithms, probably the most important
other algorithm is the algorithm of Swendsen and Wang~(1987).  In fact this
algorithm is very similar to the Wolff algorithm and Wolff took the idea
for his algorithm directly from it.  (Swendsen and Wang took the idea in
turn from the work of Fortuin and Kasteleyn~(1972) and Sweeny~(1983).)
Actually, we have seen the central idea behind the Swendsen-Wang algorithm
already.  In Section~\sref{zcpu} we considered an alternative
implementation of the Wolff algorithm in which the entire lattice of spins
is divided up into clusters by making ``links'' with probability $P_{\rm
  add} = 1 - \e^{-2\beta J}$ between similarly-oriented neighbouring spins.
We then imagined choosing a single spin from the lattice and flipping over
the whole of the cluster to which it belongs.  This procedure is clearly
equivalent to the Wolff algorithm, although in practice it would be an
inefficient way of carrying it out.

The Swendsen-Wang algorithm divides the entire lattice into clusters in
exactly the same way, with this same probability $P_{\rm add}$ of making a
link.  But then, instead of flipping just one cluster, each cluster is
independently flipped with probability $\frac12$.  We notice the following
facts about this algorithm:
\begin{enumerate}
\setlength{\itemsep}{10pt}
\item The algorithm satisfies the condition of detailed balance.  The proof
  of this fact is exactly the same as it was for the Wolff algorithm.  If
  the number of links broken and made in performing a move are $m$ and $n$
  respectively (and the reverse for the reverse move), then the energy
  change for the move is $2J(m-n)$ (or $2J(n-m)$ for the reverse move).
  The selection probabilities for choosing a particular set of links
  differs between forward and reverse moves only at the places where bonds
  are made or broken, and so the ratio of the two selection probabilities
  is $(1 - P_{\rm add})^{m-n}$, just as it was before.  By choosing $P_{\rm
    add} = 1 - \e^{-2\beta J}$, we then ensure, just as before, that the
  acceptance probability is independent of $m$ and $n$ and everything else,
  so any choice which makes it the same in each direction, such as flipping
  all clusters with probability $\frac12$ will make the algorithm correct.
  Notice however that other choices would also work.  It doesn't matter how
  we choose to flip the clusters, though the choice made here is good
  because it minimizes the correlation between the direction of a cluster
  before an after a move, the new direction being chosen completely at
  random, regardless of the old one.
\item The algorithm updates the entire lattice on each move.  In measuring
  correlation times for this algorithm, one should therefore measure them
  simply in numbers of Monte Carlo steps, and not steps per site as with
  the Metropolis algorithm.  (In fact, on average, only half the spins get
  flipped on each move, but the number flipped scales like the size of the
  system, which is the important point.)
\item The Swendsen-Wang algorithm is essentially the same as the Wolff
  algorithm for low temperatures.  Well below $T_c$, one of the clusters
  chosen by the algorithm will be a percolating cluster, and the rest will
  correspond to the ``excitations'' discussed in Section~\sref{properties},
  which will be small.  Ignoring these small clusters then, the
  Swendsen-Wang algorithm will tend to turn over the percolating backbone
  of the lattice on average every two steps (rather than every step as in
  the Wolff algorithm---see Figure~\fref{wolffsnaps}), but otherwise the
  two will behave almost identically.  Thus, as with the Wolff algorithm,
  we can expect the performance of the Swendsen-Wang algorithm to be
  similar to that of the Metropolis algorithm at low $T$, though probably a
  little slower on average due to the complexity of the algorithm.
\item At high temperatures, the Swendsen-Wang algorithm tends to divide the
  lattice into very small clusters because $P_{\rm add}$ becomes small.  As
  $T\to\infty$ the clusters will just be one spin each, and the algorithm
  will just change all the spins to new random values on each move.  This
  is also what the Metropolis algorithm does at high temperatures in one
  sweep of the lattice, though again the Metropolis algorithm can be
  expected to be a little more efficient in this regime, since it is a
  simpler algorithm which takes few operations on the computer to flip each
  spin.
\end{enumerate}

The combination of the last two points here implies that the only regime in
which the Swendsen-Wang algorithm can be expected to out-perform the
Metropolis algorithm is the one close to the critical temperature.  The
best measurement of the dynamic exponent of the algorithm is that of
Coddington and Baillie~(1992), who found $z=0.25\pm0.01$ in two dimensions,
which is clearly much better than the Metropolis algorithm, and is in fact
exactly the same as the result for the Wolff algorithm.  So the
Swendsen-Wang algorithm is a pretty good algorithm for investigating the 2D
Ising model close to its critical point.  However, as Table~\tref{zval}
shows, for higher dimensions the Swendsen-Wang algorithm has a
significantly higher dynamic exponent than the Wolff algorithm, making it
slower close to $T_c$.  The reason is that close to $T_c$ the properties of
the Ising model are dominated by the fluctuation of large clusters of spins.
As the arguments of Section~\sref{zcpu} showed, the Wolff algorithm
preferentially flips larger clusters because the chance of the seed spin
belonging to any particular cluster is proportional to the size of that
cluster.  The Swendsen-Wang algorithm on the other hand treats all clusters
equally, regardless of their size, and therefore wastes a considerable
amount of effort on small clusters which make vanishingly little
contribution to the macroscopic properties of the system for large system
sizes.  This, coupled with the fact that the Swendsen-Wang algorithm is
slightly more complicated to program than the Wolff algorithm, makes the
Wolff algorithm the algorithm of choice for most people.  It is worth
noting however, that the work of Ferrenberg, Landau, and Wong~(1992)
appears to indicate that the Wolff algorithm is unusually susceptible to
imperfections in the random number generator used to implement the
algorithm.  Although this result is probably highly sensitive to the way in
which the algorithm is coded, it may be that the Swendsen-Wang algorithm
would be a more sensible choice for those who are unsure of the quality of
their random numbers.

\begin{table}
\begin{center}
\begin{tabular}{|c|c|c|c|}
\hline
\hline
dimension $d$ & Metropolis & Wolff & Swendsen-Wang \\ 
\hline
2 & $2.167\pm0.001$ & $0.25\pm0.01$ & $0.25\pm0.01$ \\
3 & $2.02\pm0.02$   & $0.33\pm0.01$ & $0.54\pm0.02$ \\
4 & --              & $0.25\pm0.01$ & $0.86\pm0.02$ \\
\hline
\hline
\end{tabular}
\end{center}
\tcapt{Comparison of the dynamic exponent $z$ for the Metropolis, Wolff
and Swendsen-Wang algorithms in various numbers of dimensions.  Figures are
taken from Coddington and Baillie~(1992), Matz, Hunter and Jan~(1994), and
Nightingale and Bl\"ote~(1996).  To our knowledge, the dynamic exponent of
the Metropolis algorithm has not been measured in four dimensions.}
\label{zval}
\end{table}

\subsection{Niedermayer's algorithm}
\label{niedermayer}
Another variation on the general cluster algorithm theme was proposed by
Niedermayer~(1988).  His suggestion is really just an extension of the
ideas used in the Wolff and Swendsen-Wang algorithms.  In fact,
Niedermayer's methods are very general and can be applied to all sorts of
models including glassy spin models.  Here we will just consider their
application to the ordinary Ising model.

Niedermayer pointed at that it is not necessary to constrain the ``links''
with which we make clusters to be only between spins which are pointing in
the same direction.  In general, we can define two different probabilities
for putting links between sites---one for parallel spins and one for
anti-parallel ones.  The way Niedermayer expressed it, he considered the
energy contribution $E_{ij}$ that a pair of spins $i$ and $j$ makes to the
Hamiltonian.  In the case of the Ising model, for example,
\begin{equation}
E_{ij} = -J s_i s_j.
\label{eij}
\end{equation}
He then wrote the probability for making a link between two neighbouring
spins as a function of this energy $P_{\rm add}(E_{ij})$.  In the Ising
model $E_{ij}$ can only take two values $\pm J$, so the function $P_{\rm
  add}(E_{ij})$ only needs to be defined at these points, but for some of
the more general models Niedermayer considered it needs to be defined
elsewhere as well.  Clearly, if for the Ising model we make $P_{\rm
  add}(-J) = 1 - \e^{-2\beta J}$ and $P_{\rm add}(J) = 0$, then we recover
the Wolff algorithm or the Swendsen-Wang algorithm, depending on whether we
flip only a single cluster on each move, or many clusters over the entire
lattice---Niedermayer's formalism is applicable in either case.  To be
concrete about things, let us look at the case of the single-cluster,
Wolff-type version of the algorithm.

\begin{figure}
\smallfigure{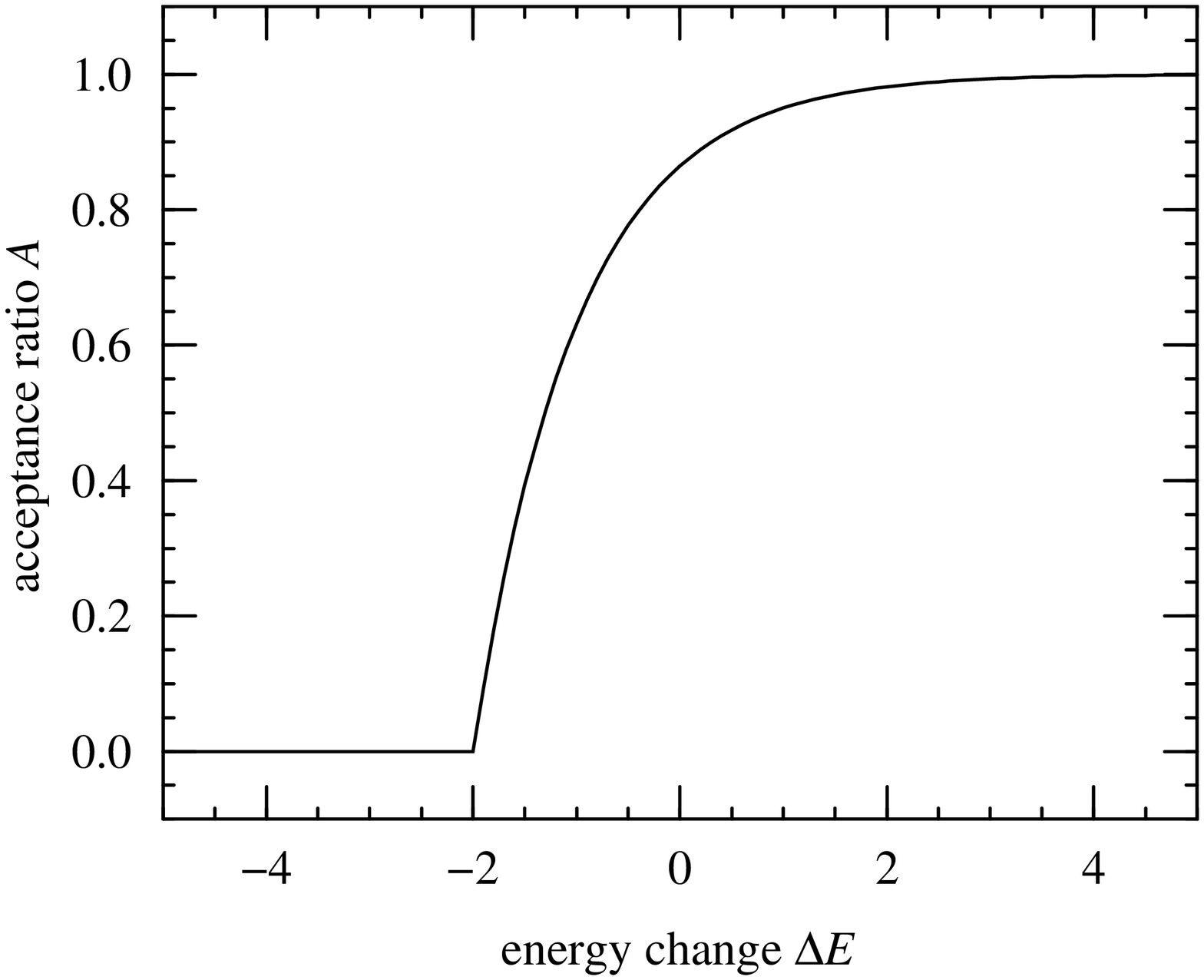}
\capt{The acceptance ratio for the Niedermayer algorithm,
  Equation~\eref{eqniedermayer}, with $\beta=1$ and $E_0=-2$ as a function
  of energy.}
\label{niedaccept}
\end{figure}

Let us apply the condition of detailed balance to the algorithm.  Consider,
as we did in the case of the Wolff algorithm, two states of our system
which differ by the flipping of a single cluster.  (You can look again at
Figure~\fref{frames} if you like, but bear in mind that, since we are now
allowing links between anti-parallel spins, not all the spins in the
cluster need be pointing in the same direction.)  As before, the
probability of forming the cluster itself is exactly the same in the
forward and reverse directions, except for the contributions which come
from the borders.  At the borders, there are some pairs or spins which are
parallel and some which are anti-parallel.  Suppose that in the forward
direction there are $m$ pairs of parallel spins at the border---these
correspond to bonds which have to be broken in flipping the cluster---and
$n$ pairs which are anti-parallel---bonds which will be made when we flip.
By definition no links are made between any of these border pairs, and the
probability of that happening is $[1 - P_{\rm add}(J)]^m [1 - P_{\rm
  add}(-J)]^n$.  In the reverse direction the corresponding probability is
$[1 - P_{\rm add}(J)]^n [1 - P_{\rm add}(-J)]^m$.  Just as in the Wolff
case, the energy cost of flipping the cluster from state $\mu$ to state
$\nu$ is
\begin{equation}
E_\nu - E_\mu = 2J(m-n).
\end{equation}
Thus, the appropriate generalization of the acceptance ratio relation,
Equation~\eref{wolffaccept}, is
\begin{equation}
{A(\mu\to\nu)\over A(\nu\to\mu)} = [\e^{2\beta J}
{1 - P_{\rm add}(J)\over1 - P_{\rm add}(-J)}]^{n-m}.
\label{eqniedaccept}
\end{equation}
Any choice of acceptance ratios $A(\mu\to\nu)$ and $A(\nu\to\mu)$ which
satisfies this relation will satisfy detailed balance.  For the Wolff
choice of $P_{\rm add}$ we get acceptance ratios which are always unity,
but Niedermayer pointed out that there are other ways to achieve this.  In
fact, all we need to do is choose $P_{\rm add}$ to satisfy
\begin{equation}
{1 - P_{\rm add}(J)\over1 - P_{\rm add}(-J)} = \e^{-2\beta J}
\end{equation}
and we will get acceptance ratios which are always one.  Niedermayer's
solution to this equation was $P_{\rm add}(E) = 1 - \exp [\beta(E_0 - E)]$
where $E_0$ is a free parameter whose value we can choose as we like.
Notice however that this quantity is supposed to be a probability for
making a bond, so it is not allowed to be less than zero.  Thus the best
expression we can write for the probability $P_{\rm add}^{(ij)}$ of adding
a link between sites $i$ and $j$ is
\begin{equation}
P_{\rm add}(E_{ij}) = \Biggl\lbrace \begin{array}{ll}
                        1 - \e^{\beta(E_0 - E_{ij})}\qquad &
                        \mbox{if $E_{ij}>E_0$}\\
                        0 & \mbox{otherwise.}
                        \end{array}
\label{eqniedermayer}
\end{equation}
And this defines Niedermayer's algorithm.  In Figure~\fref{niedaccept},
we have plotted $P_{\rm add}$ as a function of $E$ for one particular
choice of the constant $E_0$.

Notice the following things about this algorithm:
\begin{enumerate}
\setlength{\itemsep}{10pt}
\item As long as all $E_{ij}$ on the lattice are greater than or equal to
  $E_0$, the right-hand side of Equation~\eref{eqniedaccept} is always
  unity, so the two acceptance ratios can be chosen to be one for every
  move.  Since we are at liberty to choose $E_0$ however we like, we can
  always choose it to satisfy this condition by making it less than or
  equal to the smallest value that $E_{ij}$ can take, which is $-J$ in the
  case of the Ising model.  This gives us a whole spectrum of Wolff-type
  algorithms for various values $E_0\le -J$, which all have acceptance
  ratios of one.  As $E_0$ gets more and more negative, the probabilities
  $P_{\rm add}(\pm J)$ tend closer and closer to one, making the clusters
  formed larger and larger.  This gives us a way of controlling the sizes
  of the clusters formed in our algorithm, all the way up to clusters which
  encompass (almost) every spin on the lattice at every move.
\item If we choose $E_0$ be greater than the smallest possible value of
  $E_{ij}$ (which is $-J$ in the Ising case) then the right-hand side of
  Equation~\eref{eqniedaccept} is no longer equal to one, and we can no
  longer choose the acceptance ratios to be unity.  Instead we have
\begin{equation}
{A(\mu\to\nu)\over A(\nu\to\mu)} = [\e^{2\beta J} \e^{\beta(E_0 -
J)}]^{n-m} = [\e^{2\beta(E_0 + J)}]^{n-m}.
\end{equation}
Just as with the Metropolis algorithm, the optimal choice of the two
acceptance ratios is then to make the larger of the two equal to one, and
choose the smaller to satisfy this equation.  If we do this, we again
achieve detailed balance, and we now have an algorithm which, as $E_0$ is
made larger and larger, produces smaller and smaller clusters, though it
does so at the expense of an exponentially decreasing acceptance ratio for
cluster moves which increase the energy of the system.
\item If $E_0$ is chosen to be larger than the largest possible value of
$E_{ij}$, which is $+J$, then $P_{\rm add} = 0$ for all pairs of spins
$i$, $j$, so every cluster formed has only one spin in it and the
acceptance ratios are given by
\begin{equation}
{A(\mu\to\nu)\over A(\nu\to\mu)} = [\e^{2\beta J}]^{n-m}.
\end{equation}
Bearing in mind that $m$ is the number of neighbours of this single spin
which are pointing in the same direction as it, and $n$ is the number which
point in the opposite direction, we can see that this is exactly the same
as the acceptance ratio for the Metropolis algorithm.
\end{enumerate}

Thus we see that by varying the parameter $E_0$, Niedermayer's cluster
algorithm includes as special cases both the Metropolis algorithm and the
Wolff algorithm, and interpolates smoothly from one to the other and
beyond, varying the average cluster size from one spin all the way up to
the entire lattice.  The trouble with the algorithm is that no one really
knows what value one should choose for $E_0$.  Niedermayer himself
conjectured that the Wolff choice $E_0 = -J$ might not give the optimal
correlation time and that shorter ones could be achieved by making other
choices.  He gave preliminary evidence in his 1988 paper that, in some
cases at least, a larger value of $E_0$ than this (i.e.,~a value which
produces smaller clusters on average, at the expense of a lowered
acceptance probability) gives a shorter correlation time.  However, to our
knowledge no one has performed an extensive study of the dynamic exponent
of the algorithm as a function of $E_0$, so for the moment at least, the
algorithm remains an interesting extension of the Wolff idea which has yet
to find use in any large-scale simulations.

\subsection{The limited cluster algorithm}
\label{lcf}
An alternative technique for varying the sizes of the clusters flipped by
the Wolff algorithm has been studied by Barkema and Marko~(1993), who
proposed a general method for placing constraints on the clusters generated
in the Wolff algorithm.  In their algorithm, cluster growth proceeds just
as in the normal Wolff algorithm, except when the addition of a site would
violate some condition which we wish to preserve.  For example, we might
want to place a limit on the total number of spins in the cluster, or limit
the radius of the cluster around the initial seed spin.  If the addition of
a certain spin would violate such a condition, we simply make the
probability of adding it to the cluster zero, instead of the usual $P_{\rm
  add}$.  On its own, this procedure would violate detailed balance since
it alters the selection probability for the move $g(\mu\to\nu)$ away from
its carefully chosen Wolff value.  However, the requirement of detailed
balance only places a condition on the product of the selection probability
for a move and the acceptance ratio.  Thus it is possible to compensate for
the change in $g(\mu\to\nu)$ by making an opposing change to the acceptance
ratio $A(\mu\to\nu)$.  By doing this, detailed balance is restored and the
desired condition is met, albeit at the expense of changing the acceptance
ratio from the optimal value of 1 which it has in the Wolff algorithm.

In detail the method works like this.  Consider the two states $\mu$ and
$\nu$, and suppose that for a move which takes us from $\mu$ to $\nu$ there
are $m$ bonds which have to be broken in order to flip the cluster.  These
broken bonds represent correctly-oriented spins which are not added to the
cluster by the algorithm.  For most of these broken bonds, the probability
of not adding such a spin is $1-P_{\rm add}$ as in the usual Wolff
algorithm, but there will now be some number $m_0$ that were rejected
because of the applied constraint and thus had no chance of being added.
Thus the probability of not adding all the broken bonds, which is
proportional to the selection probability $g(\mu\to\nu)$ for the forward
move, is $(1 - P_{\rm add})^{m-m_0}$.  If there are $n$ bonds which are
broken in making the reverse move from $\nu$ to $\mu$, $n_0$ of which had
no chance of being added since they violated our constraint, then the
selection probability $g(\nu\to\mu)$ will be proportional to $(1 - P_{\rm
  add})^{n-n_0}$.  The ratio of the forward and backward selection
probabilities is thus the same as in the Wolff algorithm, except for a
factor of $(1 - P_{\rm add})^{m_0-n_0}$.  In order to restore detailed
balance, we compensate for this factor by introducing an acceptance ratio
of
\begin{equation}
A(\mu\to\nu) = \Biggl\lbrace \begin{array}{ll}
                \e^{-2\beta J (m_0-n_0)}\qquad & \mbox{if $m_o-n_0>0$}\\
                1 & \mbox{otherwise.}
                \end{array}
\end{equation}

What is the purpose of introducing constraints such as these?  In the
context of the normal Ising model, a constraint can be enforced to limit
the cluster to a certain region.  This allows us to divide the lattice into
a number of regions in each of which cluster flips can be performed
independently of the others.  This leads to an efficient algorithm for
simulating the model on a parallel (distributed) computer which shows
near-linear speed-up in the number of processors employed (Barkema and
MacFarland~1994).  Periodically, the division of the lattice has to be
rearranged in order to obtain ergodicity.

Another application of the idea of cluster constraints can be found in the
simulation of the conserved-order-parameter Ising model: clusters of up-
and down-pointing spins are grown with a hard constraint on their mass, and
if both clusters reach the same mass, an exchange of the spins is proposed.

Size constraints have also proved useful in the simulation of the
random-field Ising model (RFIM).  The RFIM is normally simulated using a
single-spin-flip Metropolis algorithm.  It is also possible to apply the
Wolff algorithm to the problem by growing Wolff clusters ignoring the
random fields and incorporating their energy contribution into an
acceptance ratio for the cluster flip (Dotsenko~\etal~1991).  However in
practice this algorithm works poorly in the critical region of the model
because the RFIM has a critical temperature lower than that of the normal
Ising model.  This means that the clusters grown are nearly always large
percolating clusters, which have little chance of being flipped over
because the random fields pin them in one direction or the other.  This
problem can be overcome if the size of the cluster is limited to a radius
$r$, which is chosen at random on each step from a distribution as $1/r^2$
(Newman and Barkema~1996).

\subsection{Multigrid methods}
\label{multigrid}
Also worthy of mention is a class of methods developed by
Kandel~\etal~(1989, Kandel~1991), which are referred to as multigrid
methods.  These methods are also aimed at reducing critical slowing down
and accelerating simulations at or close to the critical temperature.
Multigrid methods have not been used to a very great extent in large-scale
Monte Carlo simulations because they are considerably more complex to
program than the cluster algorithms discussed above.  However, they may yet
prove useful in some contexts because they appear to be faster than cluster
algorithms for simulating very large systems.

The fundamental idea behind these methods is the observation that, with the
divergence of the correlation length at the critical temperature, we expect
to see fluctuating domains of spins of all sizes up to the size of the
entire lattice.  The multigrid methods therefore split the CPU time of the
simulation up, spending varying amounts of time flipping blocks of spins of
various sizes.  This idea certainly has something in common with the ideas
behind the Wolff and Swendsen-Wang algorithms, but the multigrid methods
are more deliberate about flipping blocks of certain sizes, rather than
allowing the sizes to be determined by the temperature and configuration of
the lattice.  Kandel and his co-workers gave a number of different, similar
algorithms, which all fall under the umbrella of multigrid methods.  Here,
we describe one example, which is probably the simplest and most efficient
such algorithm for the Ising model.  The algorithm works by grouping the
spins on the lattice into blocks and then treating the blocks as single
spins and flipping them using a Metropolis algorithm.  In detail, what one
does is this.

In the Swendsen-Wang algorithm we made ``links'' between similarly-oriented
spins, which effectively tied those spins together into a cluster, so that
they flipped as one.  Any two spins which were not linked were free to
flip separately---there was not even a ferromagnetic interaction between
them to encourage them to point in the same direction.  In the present
multigrid method, two adjacent spins can be in three different
configurations: they can be linked as in the Swendsen-Wang case so that
they must flip together, they can have no connection between them at all so
that they can flip however they like, or they can have a normal Ising
interaction between them of strength $J$ which encourages them
energetically to point in the same direction, but does not force them to as
our links do.  By choosing one of these three states for every pair of
spins, the lattice is divided up into clusters of linked spins which either
have interactions between them, or which are free to flip however they
like.  The algorithm is contrived so that only clusters of one or two spins
are created.  No clusters larger than two spins appear.  The procedure for
dividing up the lattice goes like this.
\begin{enumerate}
\setlength{\itemsep}{10pt}
\item We take a spin on the lattice, and examine each of its neighbours in
  turn.  If a neighbour is pointing in the opposite direction to the spin,
  then we leave it alone.  In Kandel's terminology, we ``delete'' the bond
  between the two spins, so that they are free to assume the same or
  different directions with no energy cost.  If a neighbour is pointing in
  the same direction as our spin, then we make a link between the two with
  the same probability $P_{\rm add} = 1 - \e^{-2\beta J}$ as we used in the
  Wolff and Swendsen-Wang algorithms.  Kandel calls this ``freezing'' the
  bond between the spins.
\item Since we only want to create clusters of at most two spins, we stop
  looking at neighbours once we have created a link to any one of them.  In
  fact, what we do is to keep the normal Ising interactions between the
  spin and all the remaining neighbours that we have not yet looked at.  We
  do this regardless of whether they are pointing in the same direction as
  our first spin or not.  Kandel describes this as ``leaving the bonds
  active''.
\item Now we move onto another spin and do the same thing, and in this way
  cover the entire lattice.  Notice that, if we come to a spin which is
  adjacent to one we have considered before, then some of the spin's bonds
  will already have been frozen, deleted, or marked as active.  In this
  case we leave those bonds as they are, and only go to work on the others
  which have not yet been considered.  Notice also that if we come to a
  spin and it has already been linked (``frozen'') to another spin, then we
  know immediately that we need to leave the interactions on all the
  remaining bonds to that spin active.
\end{enumerate}
In this way, we decide the fate of all the spins on the lattice, dividing
them into clusters of one or two, joined by bonds which may or may not have
interactions associated with them.  Then we treat those clusters as single
spins, and we carry out the Metropolis algorithm on them, for a few sweeps
of the lattice.

But this is not the end.  Now we do the whole procedure again, treating the
clusters as spins, and joining them into bigger clusters of either one or
two elements each, using exactly the same rules as before.  (Note that the
lattice of clusters is not a regular lattice, as the original system was,
but this does not stop us from carrying out the procedure just as before.)
Then we do a few Metropolis sweeps of this coarser lattice too.  And we
keep repeating the whole thing until the size of the blocks reaches the
size of the whole lattice.  In this way, we get to flip blocks of spins of
all sizes from single spins right up to the size of the entire system.
Then we start taking the blocks apart again into the blocks that made them
up, and so forth until we get back to the lattice of single spins again.
In fact, Kandel and co-workers used a scheme where at each level in the
blocking procedure they either went towards bigger blocks (``coarsening'')
or smaller ones (``uncoarsening'') according to the following rule.  At any
particular level of the procedure we look back and see what we did the
previous times we got to this level.  If we coarsened the lattice the
previous two times we got to this point, then on the third time, we
uncoarsen.  This choice has the effect of biasing the algorithm towards
working more at the long length scales (bigger, coarser blocks).

Well, perhaps you can see why the complexity of this algorithm has put
people off using it.  The proof that the algorithm satisfies detailed
balance is even more involved, and, since you're probably not dying to hear
about it right now, we'll refer you to the original paper for the details
(Kandel~\etal~(1989)).  In the same paper it is demonstrated that the
dynamic exponent for the algorithm is in the region of $0.2$ for the
two-dimensional Ising model---a value similar to that of the Wolff
algorithm.  Before we dismiss the multigrid method out of hand, however,
let us point out that the simulations do indicate that its performance is
superior to cluster algorithms for large systems.  These days, with
increasing computer power, people are pushing simulations towards larger
and larger lattices, and there may well come a point at which using a
multigrid method could win us a significant speed advantage.

\subsection{The invaded cluster algorithm}
\label{ic}
Finally, in our round-up of Monte Carlo algorithms for the Ising model, we
come to an unusual algorithm proposed by Jon Machta and co-workers called
the invaded cluster algorithm (Machta~\etal~1995).  The thing which sets
this algorithm apart from the others we have looked at so far is that it is
not a general purpose algorithm for simulating the Ising model at any
temperature.  In fact, the invaded cluster algorithm can only be used to
simulate the model at the critical point; it does not work at any other
temperature.  What's more, for a system at the critical point in the
thermodynamic limit the algorithm is just the same as the Swendsen-Wang
cluster algorithm of Section~\sref{swalg}.  So what's the point of the
algorithm?  Well, there are two points.  First, the invaded cluster
algorithm can find the critical point all on its own---we don't need to
know what the critical temperature is beforehand in order to use the
algorithm.  Starting with a system at any temperature the algorithm will
adjust its simulation temperature until it finds the critical value $T_c$.
This makes the algorithm very useful for actually measuring the critical
temperature, something which normally requires either finite-size scaling
or Monte Carlo RG.  Second, the invaded cluster algorithm equilibrates
extremely quickly.  Although the algorithm is equivalent to the
Swendsen-Wang algorithm once it reaches equilibrium at $T_c$, its behaviour
whilst getting there is very different, and in a direct comparison of
equilibration times between the two, say for systems starting at $T=0$,
there is really no contest.  For large system sizes, the invaded cluster
algorithm can reach the critical point as much as a hundred times faster
than the Swendsen-Wang algorithm.  (A comparison with the Wolff algorithm
yields similar results---the performance of the Wolff and Swendsen-Wang
algorithms is comparable.)

So, how does the invaded cluster algorithm work?  Basically, it is just a
variation of the Swendsen-Wang algorithm in which the temperature is
continually adjusted to look for the critical point.  The algorithm finds
the fraction of links at which a percolating backbone of spins first forms
across the lattice and uses this measurement to make successively better
approximations to the critical temperature at each Monte Carlo step.  In
detail, here's how it goes:
\begin{enumerate}
\setlength{\itemsep}{10pt}
\item We choose some starting configuration of the spins.  It doesn't
  matter what choice we make, but it could, for example, be the $T=0$
  state in which all spins are aligned.
\item We add links at random between pairs of similarly-oriented
  nearest-neighbour sites until one of the clusters formed reaches
  percolation.
\item Once the links are made, we flip each cluster on the lattice
  separately with probability $\frac12$, just as we do in the normal
  Swendsen-Wang algorithm.  Then the whole procedure is repeated from
  step~(ii) again.
\end{enumerate}

Why does this algorithm work?  Well, for any given configuration of the
lattice, the maximum number of links which can be made between spins is
equal to the number of pairs of similarly-oriented nearest-neighbour spins.
Let us call this number $n$.  The number of links $m$ needed to achieve the
percolation of one cluster will normally be smaller than this maximum
value.  A given step of the algorithm is equivalent to a single step of the
Swendsen-Wang algorithm with $P_{\rm add} = m/n$.  (The number of links $m$
required for percolation is fairly constant, and most of the variation in
$P_{\rm add}$ comes from $n$.)  Substituting this value into
Equation~\eref{padd} and rearranging, we can then calculate the effective
temperature of the Monte Carlo step:
\begin{equation}
kT_{\rm eff} = -{2J\over\log(1-P_{\rm add})} = -{2J\over\log(1-m/n)}.
\label{ictemp}
\end{equation}

Consider what happens if the system is below the critical temperature.  In
this case, the spins are more likely to be aligned with their neighbours
than they are at the critical temperature, and $n$ is therefore large.
This gives us a value of $T_{\rm eff}$ which is higher than $T_c$.  In
other words, when the system is below the critical temperature, the
algorithm automatically chooses a temperature $T_{\rm eff}>T_c$ for its
Swendsen-Wang procedure.  Conversely, if the system is at a temperature
above the critical point, neighbouring spins are less likely to be aligned
with one another than they are at $T_c$, giving a lower value of $n$ and a
correspondingly lower value of $T_{\rm eff}$.  So, for a state of the
system above the critical temperature, the algorithm automatically chooses
a temperature $T_{\rm eff}<T_c$ for its Swendsen-Wang procedure.

The algorithm therefore has a kind of negative feedback built into it,
which always drives the system towards the critical point, and when it
finally reaches $T_c$ it will stay there, performing Swendsen-Wang Monte
Carlo steps at the critical temperature for the rest of the simulation.
Note that we do not need to know what the critical temperature is for the
algorithm to work.  It finds $T_c$ all on its own, and for that reason the
algorithm is a good way of measuring $T_c$.  The same feedback mechanism
also drives the algorithm towards $T_c$ faster than simply performing a
string of Swendsen-Wang Monte Carlo steps exactly at $T_c$.  This point is
discussed in more detail below.

Before we review the result of simulations using the invaded cluster
algorithm, let us briefly examine its implementation.  By and large, the
algorithm is straightforward to program, but some subtlety arises when we
get to the part about testing for a percolating cluster.  Since we need to
do this every time a link is added to the lattice, it is important that we
find an efficient test.

There are various criteria one can apply to test for percolation on a
finite lattice.  Here we use one of the simplest: we assume a cluster to be
percolating when it ``wraps around'' the periodic boundary conditions on
the lattice.  In other words the cluster percolates when there is a path
from any site leading across the lattice and back to the same site which
has an integrated displacement which is non-zero.  This test can be
efficiently implemented as follows.

All sites belonging to the same cluster are organized into a tree structure
in which there is one chosen ``reference site''.  All the other sites in
the cluster are arranged in a hierarchy below this reference site.  The
immediate daughter sites of the reference site possess vectors which give
their position relative to the reference site.  Other sites are arranged as
grand-daughter sites, with vectors giving their position relative to one of
the daughter sites, and so on until every site in the cluster is accounted
for.  The vector linking any site to the reference site can be calculated
by simply ascending the tree from bottom to top, adding the vectors at each
level until the reference site is reached.

The particular way in which the sites in a cluster are arranged on the tree
is dictated by the way in which the cluster grows.  At the start of each
Monte Carlo step no links exist and each site is a reference site.  Each
time a link is added we must either join together two separate clusters to
make one larger cluster, or join together two sites which already belong to
the same cluster.  Since our criterion for percolation is the appearance of
a cluster which wraps around the boundary conditions, it follows that
percolation can only appear on moves of the latter type.  In other words we
need only check for percolation when we add a link joining two sites which
are already members of the same cluster.  The procedure is therefore as
follows.  Each time we add a link, we use the tree structures to determine
the reference site of each of the two sites involved.  These reference
sites tell us which clusters the sites belong to.  When a link is made
between two sites belonging to different clusters, we have to combine the
two clusters into one, which we do by making one of the reference sites the
daughter of the other.  When we make a link between sites in the same
cluster, we don't need to combine two clusters, but we do need to check for
percolation.  To do this we take the lattice vectors from each of the two
sites to the reference site of the cluster and subtract them.  Normally,
this procedure should result in a difference vector of length one---the
distance between two nearest-neighbour spins.  However, if the addition of
a link between these two spins causes the cluster to wrap around the
boundary conditions, the difference of the two vectors will instead be a
unit vector {\em plus\/} a vector of length the dimension of the lattice.
In this case we declare the cluster to the percolating, and stop adding
links.

\begin{figure}
\smallfigure{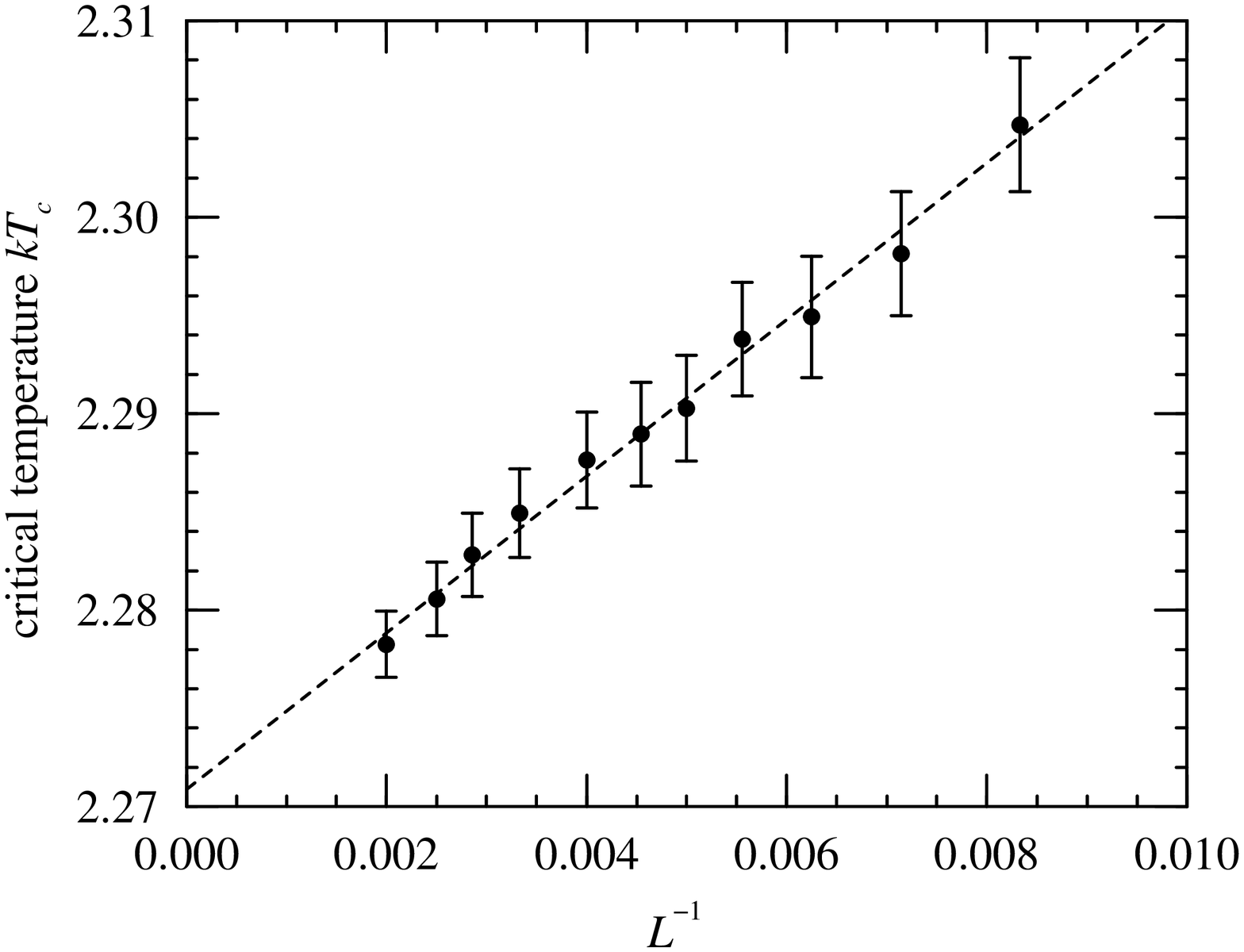} \capt{The critical temperature of the two-dimensional
  Ising model measured using the invaded cluster algorithm for systems of
  a variety of sizes from $L=120$ up to $L=500$.  Here they are plotted
  against $L^{-1}$ and the extrapolation to $L=\infty$ gives an estimate
  of $kT_c=2.271\pm0.002$ for the critical temperature in the
  thermodynamic limit.  The data are taken from Machta~\etal~(1995).}
\label{ictc}
\end{figure}

The results for the invaded cluster algorithm are impressive.
Machta~\etal~(1995) found equilibration times 20 or more times faster than
for the Swendsen-Wang algorithm at $T_c$ for the two- and three-dimensional
Ising systems they examined, and of course, the invaded cluster algorithm
allowed them to measure the value of $T_c$, which is not directly possible
with the normal Swendsen-Wang algorithm.  Figure~\fref{ictc} shows their
results for the critical temperature of a set of 2D Ising systems of
different sizes.  Because there is always a finite chance of producing a
percolating cluster on a finite lattice with any value of $P_{\rm add}$ no
matter how small (as long as it's not zero), the average estimate of $T_c$
which the algorithm makes on a finite system tends to be a little high.
However, we can easily extrapolate to the limit $L=\infty$ using
finite-size scaling.  In this case we do this by plotting the measured
$T_c$ as a function of $L^{-1}$ and extrapolating to the $L^{-1}=0$ axis.
The result is $kT_c=2.271\pm0.002$, which compares favourably with the
known exact result of $kT_c=2.269$, especially given the small amount of
CPU time taken by the simulation.  (The runs were $10\,000$ Monte Carlo
steps, for each system size.)

The invaded cluster algorithm can also be used to measure other quantities
at the critical temperature---magnetization for example, or internal
energy.  However, a word of caution is in order here.  For lattices of
finite size, which of course includes all the lattices in our Monte Carlo
simulations, the invaded cluster algorithm does not sample the Boltzmann
distribution exactly.  In particular, the fluctuations in quantities
measured using the algorithm are different from those you would get in the
Boltzmann distribution.  To see this, consider what happens once the
algorithm has equilibrated to the critical temperature.  At this point, as
we argued before, it should stop changing the temperature and just become
equivalent to the Swendsen-Wang algorithm at $T_c$, which, as we know,
certainly samples the Boltzmann distribution correctly.  However, in actual
fact, because the lattice is finite, statistical variations in the order in
which we generate the links on the lattice will give rise to variations in
the temperature $T$ of successive steps in the simulation.  The negative
feedback effect that we described above will ensure that $T$ always remains
close to $T_c$, but the size of fluctuations is very sensitive to small
changes in temperature near $T_c$ and as a result the measured
fluctuations are not a good approximation to those of the true Boltzmann
distribution.  Thus the invaded cluster algorithm is not suitable for
measuring, for instance, the magnetic susceptibility $\chi$ or the specific
heat $C$ of the Ising model at $T_c$, both of which are determined by
measuring fluctuations.  On the other hand, one could use the algorithm to
determine the value of $T_c$, and then use the normal Wolff or
Swendsen-Wang algorithm to perform a simulation at that temperature to
measure $\chi$ or $C$.

\section{Many-valued and continuous spin models}
\label{pottsetc}
Cluster Monte Carlo algorithms are not only applicable to the Ising model.
They also work for other types of spin models.  In this section we consider
how they can be modified to work with Potts models and continuous spin
models.

\subsection{Potts models}
\label{potts}
Potts models suffer from critical slowing down in just the same way as
the Ising model.  All of the algorithms discussed in this chapter can
be generalized to Potts models.  Here we discuss the example of the
Wolff algorithm, whose appropriate generalization is as follows:
\begin{enumerate}
\setlength{\itemsep}{10pt}
\item Choose a seed spin at random from the lattice.
\item Look in turn at each of the neighbours of that spin.  If they have
  the same value as the seed spin, add them to the cluster with probability
  $P_{\rm add} = 1 - \e^{-\beta J}$.  (Note that the 2 has vanished from
  the exponent; the 2-state Potts model, for example, is equivalent to
  the Ising model except for a factor two in the interaction constant $J$.)
\item For each spin that was added in the last step, examine each of {\em
    its\/} neighbours to find which ones, if any, have the same value and
  add each of them to the cluster with the same probability $P_{\rm add}$.
  (As with the Ising model, we notice that some of the neighbours may
  already be members of the cluster, in which case you don't have to
  consider adding them again.  Also, some of the spins may have been
  considered for addition before, as neighbours of other spins in the
  cluster, but rejected.  In this case, they get another chance to be added
  to the cluster on this step.)  This step is repeated as many times as
  necessary until there are no spins left in the cluster whose neighbours
  have not been considered for inclusion in the cluster.
\item Choose at random a new value for the spins in the cluster, different
  from the present value, and set all the spins to that new value.
\end{enumerate}

The proof that this algorithm satisfies detailed balance is exactly the
same as it was for the Ising model.  If we consider two states $\mu$ and
$\nu$ of the system which differ by the changing of just one cluster, then
the ratio $g(\mu\to\nu)/g(\nu\to\mu)$ of the selection probabilities for
the moves between these states depends only the number of bonds broken $m$
and the number made $n$ around the edges of the cluster.  This gives us an
equation of detailed balance which reads
\begin{equation}
{g(\mu\to\nu) A(\mu\to\nu)\over g(\nu\to\mu) A(\nu\to\mu)} =
(1 - P_{\rm add})^{m-n} {A(\mu\to\nu)\over A(\nu\to\mu)} =
\e^{-\beta(E_\nu - E_\mu)},
\end{equation}
just as in the Ising case.  The change in energy is also given by the same
expression as before, except for a factor of two:
\begin{equation}
E_\nu - E_\mu = J(m-n),
\end{equation}
and so the ratio of the acceptance ratios $A(\mu\to\nu)$ and $A(\nu\to\mu)$
for the two moves is
\begin{equation}
{A(\mu\to\nu)\over A(\nu\to\mu)} = [\e^{\beta J} (1 - P_{\rm add})]^{n-m}.
\label{contacc}
\end{equation}
For the choice of $P_{\rm add}$ given above, this is just equal to one.
Equation~\eref{contacc} is then satisfied by making the acceptance ratios
equal to 1, and with this choice the algorithm satisfies detailed balance.

As in the case of the Ising model, the Wolff algorithm gives an impressive
improvement in performance near to the critical temperature.  Arguments
similar to those of Bausch~\etal~(1981) indicate that the dynamic exponent
of the Metropolis algorithm for a Potts model should have a lower bound of
2.  By contrast, Baillie and Coddington~(1991) have measured a dynamic
exponent of $z = 0.60\pm0.02$ for the Wolff algorithm in the 2D, $q=3$
case.  As before, the single-spin-flip algorithms come into their own well
away from the critical point, because critical slowing down ceases to be a
problem and the relative simplicity of these algorithms over the Wolff
algorithm tends to give them the edge.  Ones choice of algorithm should
therefore (as always) depend on exactly which properties of the model one
wants to investigate, but the Wolff algorithm is definitely a good choice
for examining critical properties.

The Swendsen-Wang algorithm can also be generalized for use with Potts
models in a very simple fashion, as can all of the other algorithms
described in Section~\sref{further}.

\subsection{Continuous spin models}
\label{continuous}
Cluster algorithms can also be generalized to models with continuous spins.
Let us take the XY model as our example.  A version of the Wolff algorithm
for this model was given by Wolff in his original paper in 1989.  It also
works for similar models in higher dimensions.  The idea is a simple one:
one chooses at random a seed spin to start a cluster and a direction vector
$\hat{\bf n}$.\footnote{In two dimensions this easy---we simply choose an
  angle between 0 and $2\pi$ to represent the direction of the vector.  In
  three dimensions we need to choose both a $\phi$ and a $\theta$.  $\phi$
  is again uniformly distributed between 0 and $2\pi$, and $\theta$ is
  given by
  \begin{displaymath}
  \theta = \cos^{-1} (1-2r),
  \end{displaymath}
  where $r$ is random real number uniformly distributed between zero and
  one.} Then we treat the components $\hat{\bf n}\cdot{\bf s}_i$ of the spins
in that direction roughly in the same way as we did the spins in the Ising
model.  A neighbour of the seed spin whose component in this direction has
the same sign as that of the seed spin can be added to cluster by making a
link between it and the seed spin with some probability $P_{\rm add}$.  If
the components point in opposite directions then the spin is not added to
the cluster.  When the complete cluster has been built, it is ``flipped''
by reflecting all the spins in the plane perpendicular to $\hat{\bf n}$.

The only complicating factor is that, in order to satisfy detailed balance,
the expression for $P_{\rm add}$ has to depend on the values of the spins
which are joined by links thus:
\begin{equation}
P_{\rm add}({\bf s}_i,{\bf s}_j) = 1 -
\exp[-2\beta(\hat{\bf n}\cdot{\bf s}_i)(\hat{\bf n}\cdot{\bf s}_j)].
\end{equation}
Readers may like to demonstrate for themselves that, with this choice, the
ratio of the selection probabilities $g(\mu\to\nu)$ and $g(\nu\to\mu)$ is
equal to $\e^{-\beta\Delta E}$, where $\Delta E$ is the change in energy in
going from a state $\mu$ to a state $\nu$ by flipping a single cluster.
Thus, detailed balance is obeyed as in the Ising case by an algorithm for
which the acceptance probability for the cluster flip is 1.  This algorithm
has been used, for example, by Gottlob and Hasenbusch~(1993) to perform
extensive studies of the critical properties of the Heisenberg model.

Similar generalizations to continuous spins are possible for all the
algorithms discussed in Section~\sref{further}.

\section{Conclusions}
\label{conclusions}
We have discussed a number of recently developed cluster algorithms for the
simulation of classical spin systems, including: the Wolff algorithm, which
flips a single cluster of spins at each update step and almost completely
eliminates critical slowing down from the simulation of the Ising model;
the Swendsen-Wang algorithm, which updates the entire lattice at each step
and has performance similar to, though not quite as good as the Wolff
algorithm; Neidermayer's algorithm, a variation of the Wolff algorithm
which allows one to tune the size of clusters flipped; the limited cluster
algorithm, another technique for controlling the sizes of clusters;
multigrid methods, which may be even more efficient close to criticality
than the Wolff algorithm, although they achieve this at the expense of
considerable programming complexity; and the invaded cluster algorithm,
which self-organizes to the critical point of the system studied and
permits accurate measurements of $T_c$ to be made with little computational
effort.  We have also discussed briefly how algorithms such as these can be
generalized to Potts models and continuous spin models.

\section*{Acknowledgements}
The authors would like to thank Eytan Domany, Jon Machta and Alan Sokal for
useful discussions about cluster algorithms, and Daniel Kandel for
supplying a copy of his Ph.D. thesis.

\vspace{0.5in}

\section*{References}
\smallskip\baselineskip=15pt

\begin{list}{}{\leftmargin=2em \itemindent=-\leftmargin%
\itemsep=5pt \parsep=0pt}

\item {\frenchspacing Baillie, C. F. and Coddington, P. D. 1991
    Phys. Rev. B {\bf43}, 10617.}

\item {\frenchspacing Barkema, G. T. and MacFarland, T. 1994 Phys. Rev. E
    {\bf50}, 1623.}

\item {\frenchspacing Barkema, G. T. and Marko, J. F. 1993
    Phys. Rev. Lett. {\bf71}, 2070.}

\item {\frenchspacing Bausch, R., Dohm, V., Janssen, H. K. and Zia, R. K.
    P.  1981 Phys. Rev. Lett. {\bf47}, 1837.}

\item {\frenchspacing Coddington, P. D. and Baillie, C. F. 1992
    Phys. Rev. Lett. {\bf68}, 962.}

\item {\frenchspacing Dotsenko, Vl. S., Selke, W. and Talapov, A. L. 1991
    Physica A {\bf170}, 278.}

\item {\frenchspacing Ferrenberg, A. M., Landau, D. P. and  Wong,
    Y. J. 1992 Phys. Rev. Lett. {\bf69}, 3382.}

\item {\frenchspacing Fortuin, C. M. and Kasteleyn 1972 Physica {\bf57},
    536.}

\item {\frenchspacing Gottlob, A. P. and Hasenbusch, M. 1993 Physica A
    {\bf201}, 593.}

\item {\frenchspacing Kandel, D. 1991 Ph.D. Thesis, Weizmann Institute of
    Science, Rehovot, Israel.}

\item {\frenchspacing Kandel, D., Domany, E. and Brandt, A. 1989
    Phys. Rev. B {\bf40}, 330.}

\item {\frenchspacing Machta, J., Choi, Y. S., A. Lucke, Schweizer, T. and
    Chayes, L. V. 1995 Phys. Rev. Lett. {\bf75}, 2792.}

\item {\frenchspacing Matz, R., Hunter, D. L. and Jan, N. 1994
    J. Stat. Phys. {\bf74}, 903.}

\item {\frenchspacing Newman, M. E. J. and Barkema, G. T. 1996 Phys. Rev. E
    {\bf53}, 393.}

\item {\frenchspacing Niedermayer, F. 1988 Phys. Rev. Lett. {\bf61}, 2026.}

\item {\frenchspacing Nightingale, M. P. and Bl\"ote, H. W. J. 1996
    Phys. Rev. Lett. {\bf76}, 4548.}

\item {\frenchspacing Onsager, L. 1944 Phys. Rev. {\bf65}, 117.}

\item {\frenchspacing Sweeny, M. 1983 Phys. Rev. B {\bf27}, 4445.}

\item {\frenchspacing Swendsen, R. H. and Wang, J.-S. 1987
    Phys. Rev. Lett. {\bf58}, 86.}

\item {\frenchspacing Wolff, U. 1989 Phys. Rev. Lett. {\bf62}, 361 (1989).}

\end{list}

\end{document}